\documentclass[12pt]{article}
\usepackage[a4paper]{geometry}
\usepackage[italian,english]{babel}

\usepackage{amsmath}
\usepackage{amsfonts}
\usepackage{amstext}
\usepackage{amssymb}
\usepackage{amsthm}
\usepackage{amscd}

\usepackage{hyperref}
\hypersetup{colorlinks,
linkcolor=myrefcolor,
citecolor=mycitecolor,
urlcolor=myurlcolor}
\usepackage[capitalize]{cleveref}
\usepackage{caption}
\usepackage{etaremune}
\usepackage{bibentry}

\usepackage{xcolor}
\definecolor{myurlcolor}{rgb}{0,0,0.4}
\definecolor{mycitecolor}{rgb}{0,0.5,0}
\definecolor{myrefcolor}{rgb}{0.5,0,0}
\usepackage{graphicx}
\usepackage{tikz}
\usepackage{tikz-cd}
\usepackage{mathrsfs}

\usepackage{etoolbox}
\usepackage{makeidx}
\usepackage{sectsty}
\usepackage{dsfont}
\usepackage{enumitem} 
\usepackage[]{latexsym}
\usepackage{braket}
\usepackage{caption}
\usepackage[utf8]{inputenx}
\usepackage[T1]{fontenc}
\usepackage{lmodern}
\usepackage{textcomp}
\usepackage{microtype}
\usepackage{totcount}
\usepackage{blindtext}
\newtheorem*{proof*}{Proof}

\newcommand{\be}{\begin{equation}}
\newcommand{\ee}{\end{equation}}
\newcommand{\bea}{\begin{eqnarray}}
\newcommand{\eea}{\end{eqnarray}}

\title{Schwinger’s Picture of Quantum Mechanics I: Groupoids}

\author{F. M. Ciaglia$^{1,5}$  \href{https://orcid.org/0000-0002-8987-1181}{\includegraphics[scale=0.7]{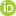}}, A. Ibort$^{2,3,6}$\href{https://orcid.org/0000-0002-0580-5858}{\includegraphics[scale=0.7]{ORCID.png}}, G. Marmo$^{4,7}$\href{https://orcid.org/0000-0003-2662-2193}{\includegraphics[scale=0.7]{ORCID.png}}\\
\footnotesize{$^{1}$\textit{ Max Planck Institute for Mathematics in the Sciences, Leipzig, Germany}} \\
\footnotesize{$^{2}$\textit{ ICMAT, Instituto de Ciencias Matem\'{a}ticas (CSIC-UAM-UC3M-UCM)}}  \\
\footnotesize{$^{3}$\textit{ Depto. de Matem\'aticas, Univ. Carlos III de Madrid, Legan\'es, Madrid, Spain}}  \\
\footnotesize{$^{4}$\textit{ Dipartimento di Fisica ``E. Pancini'', Universit\`a di Napoli Federico II, Napoli, Italy}} \\
\footnotesize{$^{5}$\textit{ e-mail: \texttt{florio.m.ciaglia[at]gmail.com}}}, \\
\footnotesize{ $^{6}$\textit{ e-mail: \texttt{albertoi[at]math.uc3m.es}}} \\ 
\footnotesize{$^{7}$\textit{ e-mail: \texttt{marmo[at]na.infn.it}}}}

\begin{document}

\maketitle

\begin{abstract}
A new picture of Quantum Mechanics based on the theory of groupoids is presented.   This picture provides the mathematical background for Schwinger's algebra of selective measurements and helps to understand its scope and eventual applications.  In this first paper, the kinematical background is described using elementary notions from category theory, in particular the notion of 2-groupoids as well as their representations.   Some basic results are presented, and the relation with the standard Dirac-Schr\"odinger and Born-Jordan-Heisenberg pictures are succinctly discussed.
\end{abstract}

\tableofcontents 

\section{Introduction} 

%%%%%%%%%%%%%%%%

\subsection{Schwinger's algebraic formulation of Quantum Mechanics}

In an attempt to establish a ``\textit{mathematical language that constitutes a symbolic expression of the properties of microscopic measurements}'', J. Schwinger proposed a family of algebraic relations between a set of symbols representing fundamental measurement processes \cite[Chap. 1]{Sc70}.  Such symbols can't consist of the classical representation of physical quantities by numbers, and they should be represented instead in terms of abstract symbols whose composition properties are postulated on the ground of physical experiences.   A similar approach was always present in Dirac's thinking on Quantum Mechanics as it was specifically stated in his lectures at Yeshiva when he dealt with q-numbers \cite{Di66}.  One of the main observations of this paper is that Schwinger's algebra of symbolic measurements can be thought of as the algebra of a certain groupoid associated with the system.
 
In Schwinger's algebraic depiction of the kinematical background of a quantum mechanical system (see \cite[Chap. 1]{Sc70}\footnote{Throughout this paper we will refer to the original edition of Schwinger's book whose notation we keep using as it is closer to the spirit of this work.}), the state of a quantum system is established by performing on it a complete selective measurement.   A general class of symbols is introduced afterwards, i.e., those denoting compound selective measurements that change a given state in another one.    Eventually, a transformation law among generalised selective measurements was postulated in order to introduce a notion of covariance in the theory.     A brief account of this perspective was offered in \cite{Ci18}, and we will extend and complete it in the present paper.

In this paper, the algebraic description of Schwinger's picture of the kinematical background of quantum mechanics systems will be established using abstract categorical notions, leaving the discussion of dynamics to a forthcoming work \cite{Ib18}.  

We will depart from an abstract setting inspired by Schwinger's conceptualisation where the primary notions will be that of  `outcomes' of measurements; `transitions', that will refer to physically allowed  changes in the outcomes of measurements; and `transformations', that is, relations among transitions as described from different experimental settings or, more generally, transformations of allowed transitions (acquiring a dynamical perspective).  These concepts are the abstractions corresponding to Schwinger's selective measurements, compound measurements and transformation functions, respectively.   

We will argue that the mathematical structure behind Schwinger's algebraic relations encompassed by the notions listed before (outcomes, transitions and transformations) is that of a 2-groupoid, where events are its 0-cells, transitions constitute the 1-cells of the 2-groupoid (and define a groupoid structure themselves), and transformations provide the 2-cells of the structure. It will be shown that such abstract notion can be used to offer an alternative picture for the description of quantum systems that, under some conditions, is equivalent to the standard pictures of Quantum Mechanics.     

\subsection{On the many pictures of Quantum Mechanics}

As it is well known,  modern Quantum Mechanics was first formulated by Heisenberg as matrix mechanics  immediately after Schr\"odinger formulated his wave mechanics. Both pictures got a better mathematical description by Dirac \cite{Di81} and Jordan \cite{Jo34}, \cite{Bo25} with  the introduction of the theory of Hilbert spaces and the corresponding theory of transformations, a sound mathematical formulation that was provided by von Neumann \cite{Ne32}. 

In all of these pictures,  the principle of analogy with classical mechanics,  as formulated by Dirac,  played a fundamental role.
The canonical commutation relations (CCR) were thought to correspond, or to be analogous to, the Poisson Brackets on phase space.  Very soon, within the rigorous formulation of von Neumann,  domain problems were identified, showing that at least one of position or momentum observables should be represented by an unbounded operator \cite{Wi47}.  Weyl introduced an ``exponentiated form'' of the commutation relations in terms of unitary operators, i.e., a projective unitary representation of a symplectic Abelian vector group, interpreted also as a phase-space with a Poisson Bracket \cite{We27}.  A $C^*$-algebra,  a generalization of the algebraic structure emerging from Heisenberg picture,  would be obtained as the group-algebra of the Weyl operators, opening the road to the highly algebraic description of quantum systems provided by $C^*$-algebras.

Thinking of relativistic quantum mechanics,  Dirac proposed the introduction of a Lagrangian formulation for quantum dynamics.  In his own words:  ``\textit{...the Lagrangian method can easily be expressed relativistically on the account of the action function being a relativistic invariant; while the Hamiltonian method is essentially non-relativistic in form, since it marks out a particular time variable as the canonical conjugate of the Hamiltonian function.}''
This suggestion was taken up by both Feynman and Schwinger, and they developed it in different directions.
Feynman's approach culminated into the path-integral formalism, where the principle of analogy is still present.
Schwinger, however,  took a different road by introducing the measurement-algebra approach, where the analogy with classical kinematics is much less evident.
Indeed, Schwinger's formulation was written for quantum systems with a finite number of states.
In any case,  for both approaches,  the seed may be found in Dirac's paper \textit{The Lagrangian in Quantum Mechanics} \cite{Di33}.

While for the various pictures associated with the names of Heisenberg, Dirac, Jordan, Weyl, etc.,   the intervening algebraic mathematical structures are nowadays clearly identified,  it is not the case for the mathematical structure underlying Schwinger's approach.  In this paper, we would like to unveil and identify this structure while postponing a thorough analysis of its implications, most notably for field theories, to forthcoming papers.

%%%%%%%%%%%%%%%%%%%%

 \subsection{Groupoids in Physics}\label{sec:categories}
 
Groupoids are playing a more relevant role in the description of the structure of physical theories.  For instance, the use of groupoids is very convenient to describe systems with internal and external structures (see for instance \cite{We86}, \cite{La98} and references therein).   It should be remarked that a groupoid structure can be identified also in the considerations made by Dirac in the previously quoted paper  \cite{Di33}. 
Indeed, the composition law of the generating functions representing transformations  allows to define a groupoid structure for the latter.
Another instance of groupoid is provided by Ritz-Rydberg combination principle of frequencies in spectral lines as observed by Connes \cite{Co94}, where groupoids are connected with the structure of certain measurements, in this case frequencies of the
emission spectrum by atoms:

\medskip

\textit{``The set of frequencies emitted by an atom does not form a group, and it is
false that the sum of two frequencies of the spectrum is again one. What
experiments dictate is the Ritz--Rydberg combination principle which permits
indexing the spectral lines by the set $\Delta $ of all pairs $(i,j)$ of
elements of a set $I$\ of indices. The frequencies $\nu _{(ij)}$\ and \ $\nu
_{(kl)}$ only combine when $j=k$ to yield $\nu _{(il)}=\nu _{(ij)}+\nu
_{(jl)}$ (...). Due to the Ritz--Rydberg combination principle, one is not
dealing with a group of frequencies but rather with a groupoid $\Delta
=\left\{ (i,j);i,j\in I\right\} $ having the composition rule $%
(i,j)(j,k)=(i,k)$. The convolution algebra still has a meaning when one
passes from a group to a groupoid, and the convolution algebra of the
groupoid $\Delta $ is none other that the algebra of matrices since the
convolution product may be written $(ab)_{(i,k)}=\sum_{n}a_{(i,n)}b_{(n,k)}$
which is identical with the product rule of matrices."}

\medskip

It is quite convenient to think of groupoids as
codifying processes in the sense that the composition law determines the different ways one can use to pass from one base element (the objects of the groupoid) to another one\footnote{For a friendly introduction to groupoids we refer the reader to \cite{Ib13}, \cite{Ib19b}.}.   Actually, the best way to think about a groupoid
is in terms of categories or, put in a different way, a category is a broad generalization of the notion of a group(oid).

The process of abstracting properties of physical systems obtained by their observation, like the properties of measurements on microscopic systems pondered by Schwinger, is extremely useful by itself.
However, as it was pointed out by J. Baez, the mathematical language based on set theory is extremely restrictive and limited for many purposes.  
Physics dealing with processes and relations both at the classical and quantum level, is particularly bad suited to be described by set theory (see for instance \cite{Ba01}).    

On the contrary, category theory is exactly about that, the emphasis is not in the description of the elements of sets, but on the relations between objects, i.e., `morphisms'.  Therefore, `elements' in set theory correspond to `objects' of a category (and `elements' can be `sets' themselves without incurring in contradictions!) and `equations between elements' correspond to `isomorphisms between objects'.   `Sets' correspond in this categorification of abstract notions to `categories' and maps between sets to `functors' between categories.   

In particular, a representation of a given category is a functor from this category in the category whose objects and morphisms are those mathematical structures we want to use to `realise' our category, e.g., linear spaces and linear maps  in the case of linear representations. Finally `equations between functions' will correspond to `natural transformations' between functors.     

Considering functors themselves as objects and natural transformations as morphisms led to the notion of higher categories, in particular, the 2-categories and the corresponding notions of 2-groups and 2-groupoids.  These abstract notions are gaining more and more interest in the description of physical phenomena (see for instance recent applications to describe topological matter \cite{Ka17},\cite{Al17}).  Surprisingly enough, we will argue that this highly abstract notions, in particular the notion of 2-groupoid, is just what is needed to provide the formal mathematical background to the kinematics of Schwinger's algebra of selective measurements.

The structure of this paper is as follows.  First, the basic notions of categories and groupoids will be described in a succinct way.  Then, it will be sketched how the descripton of physical systems based on groupoids would lead in a natural way to the notion of a 2-groupoid, and eventually, in the finite case, it will be shown how Schwinger's algebra of selective measurements is an instance of it.   In the meantime, a sketch of a theory of representations will be developed and some fair connections with the standard descriptions of Quantum Mechanics will be outlined.

%%%%%%%%%%%%%%%%%%%
%%%%%%%%%%%%%%%%%%%

\section{Groupoids and categories}

%%%%%%%%%%%%%%%%%%%

\subsection{Categories}

Using the language of Category Theory a groupoid is a category all of whose morphisms are invertible.
Let us recall that a category \textbf{C} consists of a family of objects $x,y,\ldots$, denoted collectively as Ob(\textbf{C}), and a family of morphisms (or arrows)  $\alpha \colon x \to y$, $\beta \colon u \to v$,...  denoted collectively as Mor(\textbf{C}).   Given two objects $x,y$, the family of morphisms from $x$ to $y$ is denoted as Mor($x,y$).   The category $\mathbf{C}$ is equipped with a composition law that assigns to any pair of morphisms $\alpha\colon x \to y$ and $\beta\colon y \to z$ a morphism\footnote{This convention mimicks the notation for the composition of functions.} $\beta \circ \alpha \colon x \to z$.  The composition law is associative,  that is, $(\alpha  \circ \beta) \circ \gamma = \alpha \circ (\beta \circ \gamma)$ whenever $\alpha, \beta$ and $\gamma$ can be composed.    Finally it is assumed that there exists a family of morphisms $1_x$ such that $\alpha\circ 1_x = \alpha$ and $1_y \circ \alpha = \alpha$ for any $\alpha \colon x \to y$.    

Sometimes it would be convenient to denote the category $\mathbf{C}$ as: Mor(\textbf{C}) $\rightrightarrows$ Ob(\textbf{C}) where the double arrows denote the assignments to each morphism $\alpha \colon x \to y$ of the `source' object $x$ and the `target' object $y$ respectively.  In this sense we will denote $x = s(\alpha)$ and $y = t(\alpha)$ (it will be also denoted sometimes $y = \alpha(x)$).  Notice that the morphism $\alpha$ can be composed with the morphism $\beta$ iff $t(\alpha) = s(\beta)$.  

A morphism $\alpha\colon x \to y$ is said to be invertible if there exists $\beta\colon y \to x$ such that $\alpha\circ \beta = 1_y$ and $\beta\circ \alpha = 1_x$.  Such morphism will be called the inverse of $\alpha$ and will be denoted as $\alpha^{-1}$.   An invertible morphism will be called an isomorphism.   Notice that for any given category \textbf{C} the subcategory of invertible morphisms is a groupoid, called the groupoid of the category.
  
It will be assumed in what follows that all categories considered are small, that is, their objects and family of objects as well as the family of all morphisms are sets, and morphisms are maps among sets.   Then the previous notation for objects and morphisms coincide with the corresponding set-theoretical notions.   However it is important to bear in mind that many interesting examples, potentially relevant in the considerations on the foundations of Quantum Mechanics, could involve categories which are not small.  Consider for instance the category \textbf{Vect} of all vector spaces whose objects are complex linear spaces and morphisms are linear maps between them.     Such category is larger than the category \textbf{Sets} (whose objects are sets and morphisms are maps among sets) because given a set $S$ we may construct the complex linear space $V(S)$ freely generated by $S$, however the family of all sets is not a set, hence the family of all linear spaces is not a set and the category \textbf{Vect} is not small.  

%%%%%%%%%%%%%%%

\subsection{Groupoids}\label{sec:groupoids}

Even if the notion of groupoid is categorical, along this paper groupoids will be considered to be sets (even finite in Schwinger's conceptualisation).  Thus, in this section abstract groupoids
$\mathbf{G}$ will be briefly discussed from a set-theoretical perspective, that is, we will consider groupoids $\mathbf{G}$ whose objects form a set $\Omega$ and whose morphisms are elements $\gamma$ of the set $\mathbf{G}$.  

Two maps $s$ (`source') and $t$ (`target') will be defined from $\mathbf{G}$ onto $\Omega$, such that there is a binary operation $\circ $ (called multiplication) which is defined
for pairs $\gamma$ and $\mu$ of elements in $\mathbf{G}$ whenever $%
t(\gamma) = s(\mu)$ (then $\gamma$ and $\mu$ will be said to be composable) and the resulting element will be denoted $\mu\circ \gamma$ (notice the backwards notation for composition consistent with the convention introduced in the case of categories and the notation used for the composition of maps).  We will keep using the diagrammatic notation $\gamma \colon x \to y$  if $s(\gamma) =x$ and $t(\gamma) = y$ as in the abstract categorial setting even if $\gamma$ is not a map between sets. 

Moreover for a given groupoid $\mathbf{G}$, the maps $\{ s,t,\circ \} $ must satisfy the following axioms:

\begin{enumerate}
\item[a)] $s(\mu \circ \gamma) = s(\gamma)$, $t( \mu \circ \gamma) = t(\mu)$ for all composable $\gamma $ and $\mu$.

\item[b)] There exists $1_{s(\gamma)}$, and $1_{t(\gamma)}$ elements in $\mathbf{G}$ which are left and right unities for $\gamma$
respectively, i.e., $1_{t(\gamma )}\circ \gamma =\gamma$, $\gamma \circ 1_{s(\gamma)} =\gamma$, for all $\gamma \in \mathbf{G}$.

\item[c)] The multiplication $\circ $ is associative: if $(\gamma \circ
\mu )\circ \nu$ is defined, then $\gamma \circ \left(
\mu \circ \nu \right) $ exists and $(\gamma \circ \mu)\circ \nu = \gamma \circ \left( \mu\circ \nu \right) $.

\item[e)] Any $\gamma $ has a two-sided inverse $\gamma ^{-1},$ with $%
\gamma \circ \gamma ^{-1}=1_{t(\gamma )}$ and $\gamma ^{-1}\circ \gamma
=1_{s(\gamma )}$. The map $\mathrm{inv}%
\colon \gamma \rightarrow \gamma ^{-1}$ is an involution, that is $\left(
\gamma ^{-1}\right) ^{-1}=\gamma $.
\end{enumerate}

There is a natural equivalence relation defined in the space $\Omega$ of objects of a groupoid: $x \sim y$ \
iff there exists $\gamma \in \mathbf{G}$ such that $\gamma \colon x \to y$.   Eleements $y\in \Omega$ equivalent to $x$ form an equivalence class denoted by $\mathcal{O}_x$.   Any such equivalence class is called an orbit
of $\mathbf{G}$ and $\Omega$ is the union of all these orbits.

The set $G_x = \{ \gamma \in \mathbf{G} \mid s(\gamma )= t(\gamma ) = x \}$ is a group, called the isotropy group of $x \in
\Omega$.  Notice that the isotropy groups $G_x$, $G_y$, of objects $x, y$ in the same orbit are isomorphic (even though not canonically).

Two extreme cases regarding the space of objects of a groupoid occur when $\Omega$ is the whole groupoid $\mathbf{G}$, or $\Omega$ consists of just one point. In the first instance 
$s(\gamma ) = t(\gamma )=\gamma $ for all $\gamma$ and any $\gamma $ can
be composed only with itself yielding $\gamma \circ \gamma =\gamma$. The
orbits  are single elements $\gamma $ and the isotropy group of any $\gamma$ is trivial, containing only $1_\gamma$, the theory is rather dull and becomes just the theory of sets.
However,  in the latter situation, $\Omega = \left\{ x \right\}$, so that $\mathbf{G}$ is a group, there is only one orbit $
\left\{ x \right\} $, and the isotropy group of $x$ is $\mathbf{G}$.

A groupoid $\mathbf{G}$ is called connected (or transitive) if the map
\begin{equation}
\left( s,t \right)  \colon \mathbf{G} \rightarrow \Omega \times \Omega \, , \qquad \gamma \mapsto \left(
s\left( \gamma \right) ,t\left( \gamma \right) \right) \, , 
\end{equation}%
is onto, which is equivalent to say that $\mathbf{G}$ has only one orbit $\Omega$. Finally a groupoid $\mathbf{G}$ is called principal if $(s,t)$ is one-to-one (we will also say that $\mathbf{G}$ is the groupoid of pairs of the set $\Omega$, see the discussion below, Sect. \ref{sec:examples}). 

Note that, as a set, any groupoid is the disjoint union of groupoids $%
\mathbf{G} = \sqcup _{i}\mathbf{G}_{i}$ corresponding to the partition of $\Omega = \sqcup _{i}\mathcal{
O}_{i}$ into orbits $\mathcal{O}_{i}$.  Each $\mathbf{G}_{i}$ has only one orbit and elements in $\mathbf{G}_{i}$ cannot be composed with elements in $\mathbf{G}_{k}$ whenever $k\neq i$. 

\subsection{Two simple examples: the groupoid of pairs and the action groupoid}\label{sec:examples}

Two simple, but significative, examples of groupoids will be discussed here: the groupoid of pairs of a set and the action groupoid corresponding to the action of a group on a given space.   Of course, as it was indicated before, any group $G$ is a groupoid, actually any groupoid with just one object is a group and the isotropy group of such element is the groupoid itself.    However other extreme situation happens to be of great importance, specially groupoids such that the isotropy group $G_x$ is trivial for all objects.  This correspond to the groupoid of pairs of a set.

\subsubsection{The groupoid of pairs of a set} 
\begin{figure}[h]
  \centering
    \resizebox{13cm}{5cm}{\includegraphics{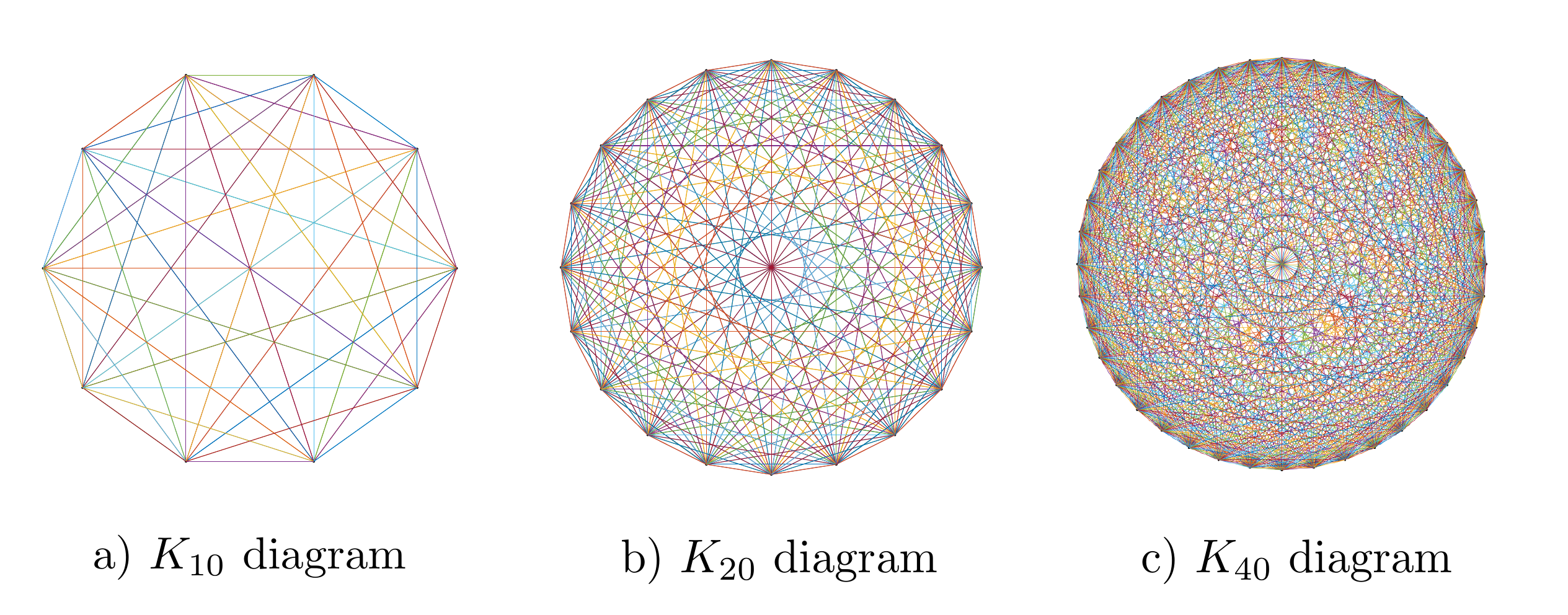}} 
    \caption{Diagrams of the groups of pairs of $n$ elements corresponding to the complete graphs $K_n$ with $n=10,20,40$.}
  \label{complete_graph}
\end{figure}

The groupoid $\boldsymbol{\Gamma} (\Omega)$ of pairs of an arbitrary set $\Omega$ is the groupoid whose objects are the elements $x$ of the set $\Omega$ and whose morphisms are pairs $(x,y)\in \Omega \times \Omega$, that is, a groupoid element $\gamma$ is just a pair $(x,y)$; $y$ will be the source of $\gamma$ and $x$ its target, i.e., any such groupoid element could be writen as $\gamma \colon y \to x$.   The composition will follow the standard usage:  $(x,y) \circ (y,z) = (x,z)$.    The unit morphisms are given by $1_x = (x,x)$ and the inverse of the morphisms $\gamma = (x,y)$ is $\gamma^{-1} = (y,x)$.  Notice that the isotropy group of any $x \in \Omega$ is the trivial group $G_x = \{ 1_x\}$.   The groupoid of pairs of a finite set $\Omega$ of $n$ elements can be drawn as the complete graph of $n$ vertices (see Fig. \ref{complete_graph}) where links represent both morphisms $(i,j)$ and $(j,i)$.

The groupoid of pairs $\boldsymbol{\Gamma} (\Omega)$ is connected as any element $y$ can be joined to any other $x$ by the morphism $(x,y)$.  

\subsubsection{The action groupoid} The action groupoid conveys globally the intuitive notion of a groupoid as acting `locally' on a set.    Thus, let $\Phi \colon \Gamma \times \Omega \to \Omega$ denote an action of the group $\Gamma$ on the set $\Omega$, that is, $\Phi(e,x) = x$, $\Phi (g , \Phi(g', x)) = \Phi (gg',x)$ for all $x \in \Omega$ and $g,g' \in \Gamma$.   As customary we will denote by $g\cdot x$ (or just $gx$) the action $\Phi (g,x)$ of the element $g$ on $x$.    We will denote by $\mathbf{G}(\Gamma,\Omega)$ the groupoid whose morphisms are pairs $(g,x) \in \Gamma \times \Omega$, the source map given by $s(g,x) = x$ and the target map, $t(g,x) = gx$.   The composition law is given by $(g',x')\circ (g,x) = (g'g,x)$ provided that $t(g,x) = gx = x' = s(g',x')$.   The unit morphisms are given by $1_x = (e,x)$ and the inverse of the morphism $\gamma = (g,x)$ is given by $\gamma^{-1} = (g^{-1}, gx)$.  

It is clear that the orbit $\mathcal{O}_x$ of the element $x\in \Omega$ is the set of elements $y$ which are targets of morphisms with source at $x$, that is $y = gx$, $g \in \Gamma$, in other words the standard notion of orbit of an element under the action or the group $\Gamma$.  

 The groupoid $\mathbf{G}(\Gamma, \Omega)$ is connected iff the action of the group is transitive.  The isotropy group of any element $x$ is the set of morphisms $\gamma \in s^{-1}(x) = t^{-1}(x) = \{ g \in G \mid x = gx \}$ which agrees with the standard notion of isotropy group.  
 
 It is remarkable that given any subset $S \subset \Omega$, the restriction of the groupoid $\mathbf{G}(\Gamma, \Omega)$ to $S$, that is the set of pairs $(g,x) \in \Gamma \times S$ such that $gx \in S$, is a subgroupoid of $\mathbf{G}(\Gamma, \Omega)$ which, in general, is not the action groupoid of any group acting on $S$.

%%%%%%%%%%%%%%%%%
%%%%%%%%%%%%%%%%%

 \subsection{The groupoid algebra and representations of finite groupoids}\label{sec:groupoid_algebra}

We illustrate the previous ideas by considering  finite groupoids and their representations.
Thus we will consider a finite groupoid $\mathbf{G}$ of order $N = |\mathbf{G}|$.   Thus we can label the morphisms of $\mathbf{G}$ as $\gamma_k$:
\begin{equation}
\mathbf{G} =\left\{ \gamma _{k}\mid k = 1\ldots, N\right\} \, .
\end{equation}
Necessarily the space of objects $\Omega$ will be finite.   Denoting by $n$ the number of elements of $\Omega$ we can label them as $\{ x_a \mid a = 1, \ldots, n \}$, $n = |\Omega|$.  

The groupoid algebra $\mathbb{C}[\mathbf{G}]$ of the finite groupoid $\mathbf{G}$ is the associative algebra generated by the elements of the groupoid with the corresponding natural composition law, that is, if $\mathbf{a} = \sum_{k} a_k \gamma_k$ and $\mathbf{b} = \sum_{l} b_l \gamma_l$, $a_k,b_l \in \mathbb{C}$, are two finite formal linear combination of elements in $\mathbf{G}$ (hereafter any summation label $j,k$,... ranges from $1$ to $N$), we define its product as:
$$
\mathbf{a} \cdot \mathbf{b} = \sum_{k,l} a_k\, b_l \, \, \delta(\gamma_k,\gamma_l)\, \, \gamma_k\circ \gamma_l \, ,
$$
with the indicator function $ \delta(\gamma_k,\gamma_l)$ defined to be $1$ if $\gamma_k,\gamma_l$ can be composed and zero otherwise.   
The product defined in this way is clearly associative because the composition law $\circ$ is associative.
The groupoid algebra $\mathbb{C}[\mathbf{G}]$ has a unit given by $1 = \sum_a 1_{x_a}$.   

Notice that the canonical basis of the groupoid algebra provided by the elements $\gamma_k$ of the groupoid allow to identified the groupoid algebra $\mathbb{C}[\mathbf{G}]$ with the algebra of complex valued functions 
$\mathcal{F}\left( \mathbf{G}\right) = \{ f \colon \mathbf{G} \to \mathbb{C}\}$ on the groupoid with the convolution product:
\begin{equation}
\left( f_{1}\ast f_{2}\right) (\gamma _{i})=\sum_{\gamma
_j \circ \gamma_k = \gamma_i} f_{1}(\gamma _{j})f_{2}(\gamma _{k}) \, ,
\end{equation}%
with $f_1, f_2 \colon \mathbf{G} \to \mathbb{C}$ any two such functions.
The identification is provided by the correspondence $\mathbb{C}[\mathbf{G}] \to \mathcal{F}(\mathbf{G})$ defined by $\mathbf{a}\mapsto f_\mathbf{a}$, with the function $f_\mathbf{a}$ defined by $f_\mathbf{a} (\gamma_k) = a_k$.  The converse map being defined as $f \mapsto \mathbf{a}_f = \sum_k f(k) \gamma_k$.  Notice that clearly:
$$
f_{\mathbf{a}} \ast f_{\mathbf{b}} = f_{\mathbf{a}\cdot \mathbf{b}} \, ,
$$
and
$$
\mathbf{a}_f \cdot \mathbf{a}_g = \mathbf{a}_{f \ast g} \, .
$$

The functions $\delta
_{\gamma _{j}},$ defined as%
\begin{equation}
\delta _{\gamma _{j}}(\gamma _{k})=\left\{
\begin{array}{c}
1\ \mathrm{if}\ \gamma _{j}=\gamma _{k} \\
0\ \ \mathrm{if}\ \gamma _{j}\neq \gamma _{k}%
\end{array}%
\right. ,
\end{equation}%
determine a basis of the groupoid algebra.  Thus for any 
$ f\in \mathcal{F}\left( \mathbf{G}\right)$, we may write:
\begin{equation}
f =\sum\limits_{k}f\left( \gamma _{k}\right) \delta
_{\gamma _{k}} \, .
\end{equation}%
Moreover:
\begin{equation}
\left( \delta _{\gamma _{j}}\ast f\right) (\gamma _{i})=\sum\limits_{\gamma _{j}\circ \gamma _{k}=\gamma _{i}} f(\gamma _{k}).
\end{equation}%
In particular $\delta _{\gamma _j}\ast \delta _{\gamma _h}$ is $1$ on $\gamma _{j}\circ \gamma _{k}$ if $\gamma _{j}$ and $\gamma _{k}$ are composable and $0$ elsewhere;  so
\begin{equation}\label{gammajk}
\delta _{\gamma _{j}}\ast \delta _{\gamma _{h}}=\delta _{\gamma _{j}\circ
\gamma _{k}} = \delta(\gamma_j, \gamma_k) \, .
\end{equation}%

The groupoid algebra $\mathbb{C}[\mathbf{G}]$  carries also an involution operator $*$ defined as $\mathbf{a}^* = \sum_k \bar{a}_k \gamma_k^{-1}$ for any $\mathbf{a} = \sum_k a_k \gamma_k$, or, in terms of the isomorphic algebra of functions:
$$
f^* = \sum_k \overline{f(\gamma_k^{-1})} \delta_{\gamma_k} \, ,
$$ 
for any $f = \sum_k f(\gamma_k) \delta_{\gamma_k}$.

A linear representation of a groupoid $\mathbf{G}$ is a functor $\rho \colon \mathbf{G} \to \mathbf{Vect}$, that is, the functor $\rho$ assigns to any object $x \in \Omega$ a linear space $\rho(x) = V_x$, and to any morphism $\gamma \colon x \to y$, a linear map $\rho(\gamma) \colon V_x \to V_y$ such that $\rho(1_x) = \mathrm{id}_{V_x}$ and $\rho(\gamma \circ \gamma' ) = \rho(\gamma) \rho(\gamma')$ for any $x \in \Omega$ and any composable pair $\gamma, \gamma'$.    Thus the notion of linear representation of groupoids extends in a natural way the theory of linear representations of groups.    

Notice that given a finite groupoid $\mathbf{G}$ there is a natural identification between linear representations $\rho$ of the groupoid and $\mathbb{C}[\mathbf{G}]$-modules.  The correspondence is established as follows.   Let $\rho$ be a representation of $\mathbf{G}$, then we define the linear space $V = \bigoplus_a V_{x_a}$ and the map $R \colon \mathbb{C}[\mathbf{G}] \to \mathrm{End}(V)$ as $R(\mathbf{a}) (v) = \sum_k a_k \rho(\gamma_k)(v)$, with $\rho(\gamma_k) (v) = \rho(\gamma_k) (v_a)$ if $\gamma_k \colon x_a \to x_b$ and $v = \oplus_a v_a$.  The map $R$ is clearly a homomorphism of algebras, hence we may consider the linear space $V$ as a (left-) $\mathbb{C}[\mathbf{G}]$-module.    

The converse of this correspondence is obtained by defining the subspaces $V_a$ of the $\mathbb{C}[\mathbf{G}]$-module $V$ by means of the projectors $P_a = R(1_{x_a})$, that is $V_a = P_a(V)$.   Then $\rho(x_a) = V_a$ and $\rho(\gamma_k) (v_a) = R(\gamma_k)(v_a)$ if $\gamma_k \colon x_a \to x_b$. 

Any finite groupoid $\mathbf{G}$ possesses two canonical representations: the fundamental and the regular representations.
We will describe briefly both of them in what follows.   

%%%%%%%%%%%%%%%%%

\subsubsection{The fundamental representation of a finite groupoid}\label{sec:fundamental}

Given the finite groupoid $\mathbf{G}$ with object space $\Omega$, we define the Hilbert space $\mathcal{H}_\Omega$ as the complex linear space generated by the elements $x \in \Omega$ with inner product:
$$
\langle \phi, \psi \rangle = \sum_{a = 1}^n \bar{\phi}_a \psi_a \, ,
$$
with $\phi = \sum_a \phi_a |x_a\rangle$, $\phi_a \in \mathbb{C}$, and where we have indicated by $| x \rangle$ the vector associated with the element $x \in \Omega$.  Notice that with this definition $\langle y, x \rangle = \delta_{xy}$ and the set of vectors $|x_a\rangle$ form an orthonormal basis of $\mathcal{H}_\Omega$.  Again, using the previous notation $\mathcal{H}_\Omega = \bigoplus _{a= 1}^n
 \mathbb{C} | x_a \rangle$.  

The fundamental representation of $\mathbf{G}$ assigns to any object $x\in \Omega$ the linear space $\pi(x) = \mathbb{C}|x \rangle$ and to any groupoid element $\gamma \colon x \to y$, the linear map $\pi(\gamma) \colon \pi(x) \to \pi(y)$, given by $\pi(\gamma) |x\rangle = |y\rangle$.  

Because of the one-to-one correspondence between linear representations of groupoids and modules, we may define the fundamental representation by the map $\pi \colon \mathbb{C}[\mathbf{G}] \to \mathrm{End}(\mathcal{H}_\Omega)$, that provides such module structure, given by:
$$
\pi(\mathbf{a})\phi = \sum_{k,b} a_k\phi_b  \, \pi(\gamma_k)|x_b \rangle \, .  
$$
Introducing the indicator symbol $\delta (\gamma_k,x_b)$ defined as 1 if $s(\gamma_k) = x_b$ and zero otherwise, we can write the previous equation as:
$$
\pi(\mathbf{a})\phi = \sum_{k,b} a_k\phi_b \, \,  \delta (\gamma_k,x_b) \,   |t(\gamma_k) \rangle \, .  
$$
Notice that the fundamental representation is a $*$-representation, that is, $\pi(f^*) = \pi(f)^\dagger$ where $\pi(f)^\dagger$ denotes the adjoint operator with respect to the inner product structure in $\mathcal{H}_\Omega$ (notice that $\langle y , \pi(\gamma) x \rangle = \langle \pi(\gamma^{-1})y , x \rangle$ if $\gamma \colon x \to y$).  

The fundamental representation allows us to introduce a natural norm on the groupoid algebra as:
\begin{equation}\label{norm}
|| f || = || \pi(f) || \, , \qquad f \in \mathbb{C}[\mathbf{G}] \, .
\end{equation}
where the norm in the r.h.s. of Eq. (\ref{norm}) is the operator norm.  Then it is trivial to check that $|| f^* f || = || f ||^2$, which means that the groupoid algebra $\mathbb{C}[\mathbf{G}]$ is a $C^*$-algebra.   The construction of a $C^*$- groupoid algebra can be done in general by selecting a family of left-Haar measures on $\mathbf{G}$ (see for instance \cite{Renault} for details).  

%%%%%%%%%%%%%%%%%

\subsubsection{The regular representation of a groupoid}
A representation $R\colon \mathcal{F}(\mathbf{G}) \to \mathrm{End}(\mathcal{F}(\mathbf{G}))$ of the groupoid algebra $\mathbb{C}[\mathbf{G}]$ (identified with $\mathcal{F}(\mathbf{G})$) on its space of functions, is obtained immediately by using the formula:
\begin{equation}
R(f) = \sum\limits_{k}f\left( \gamma _{k}\right) D_{\gamma _{k}} \, .
\label{D quant}
\end{equation}%
where 
\begin{equation}
D_\gamma (\cdot) =\delta _{\gamma }\ast \cdot \, , \label{deltagr}
\end{equation}%
because from Eq. (\ref{gammajk}), we get:
\begin{equation}
D_{\gamma _j} D_{\gamma _k} = \left\{
\begin{array}{cl}
D_{\gamma _j\circ \gamma _k} & \mathrm{\,\, if\,}\gamma _{j}\circ
\gamma _{k}\mathrm{\,\, exists,} \\ 
0 & \mathrm{\,\, if\, }\gamma _{j}\circ \gamma _{k}\mathrm{\,\,
does~not~exist.}%
\end{array}%
\right.
\end{equation}%
Notice that, consistently, we get:
\begin{eqnarray}
R(f_{1}) R(f_{2}) &=&\sum_{j,k}f_{1}(\gamma _{j})f_{2}(\gamma _{k})D_{\gamma
_j \circ \gamma _k} = \sum_{\overset{i,j,k}{{\gamma _{j}\circ \gamma }_{k}{%
=\gamma }_{i}}}f_{1}(\gamma _{j})f_{2}(\gamma _{k})D_{\gamma _i} \\
&=&\sum_{i}\left( f_{1}\ast f_{2}\right) \left( \gamma _{i}\right) D_{\gamma
_i} =R(f_{1}\ast f_{2}).  \notag
\end{eqnarray}%
In other terms, the product of operators corresponds to the convolution
product of the associated functions.  The representation $R$ will be called the (left) regular representation of the groupoid algebra (there is a similar definition of the groupoid algebra acting as a right-module on the space of functions).

Notice that because the groupoid is finite we may identify the space of functions on it with the space of square integrable functions with respect to the natural inner product defined by the standard basis $\delta_\gamma$.  In that case it is easy to check again that the regular representation is a $\ast$-representation.

%%%%%%%%%%%%%%%%%%%%

%%%%%%%%%%%%%%%%%%%%%
%%%%%%%%%%%%%%%%%%%%%

\section{2-groupoids and quantum systems}\label{sec:systems}

%%%%%%%%%%%%%%%

\subsection{The inner groupoid structure: events and transitions}\label{sec:transitions}

Given a physical system let us denote by $\mathcal{E}$ an ensemble associated with it, that is a large family of physical systems of the same type and satisfying the same specified conditions \cite{Es99}.   The elements $S$ of the ensenble $\mathcal{E}$, that is, individual systems,  are of course noninteracting.
 
It will be assumed that there is a family of observables $\mathcal{A}$ representing measurable physical quantities, and that the outcomes $a \in \mathbb{R}$ of their individuals $A \in \mathcal{A}$ can be obtained by performing physical measurements on elements $S$ of the ensemble $\mathcal{E}$ (ideally to each $A$ corresponds one particular device by means of which the measurement is made).  Any such specific measurement will be denoted as $a = \langle A:S\rangle$ (our notation) and it clearly supposes an idealized simplification of a full fledged statistical interpretation of a quantum mechanical picture.

It will be assumed that the ensemble $\mathcal{E}$ is large enough so that for any observable $A$ and for each possible outcome $a$ of the observable there are elements $S$ of the ensemble such that when $A$ is measured on them the outcome is actually $a$.  Under these conditions we will say that the ensemble $\mathcal{E}$ is \textit{sufficient} for the observable $A$\footnote{Note that we may also proceed by dictating that the observable $A$ is defined by limiting the outcomes to the actual values that can be measured over the elements of the ensemble $\mathcal{E}$, however, the natural notion of sub-ensemble leads immediately to consider maximal ensembles that will be `sufficient' in the previous sense.}.
 
The measurements we are referring to are assumed to be non-destructive, that is, the act of measuring the observable $A$ on $S$ is separated by the act of registering the outcome.
Essentially, if we think to the Stern-Gerlach experiment, we are excluding from the experimental apparatus the screen which the silver atoms hit after experiencing the magnetic field.
This instance allows us to define the notion of compatible observables as follows.
Two observables $A,B \in \mathcal{A}$ will be said to be compatible if the outcomes of their respective measurements are not affected by the outcomes of the other.   Alternatively we may say that their outcomes do not depend on the order in which they are performed.  If we denote by $\langle A,B:S\rangle$ the outcome $(a,b)$ obtained of the measurement of the ordered pair of observables $A$ and $B$ on the element $S$ of the ensemble, that is, first $B$ is measured and, after a negligible amount of time, $A$ is measured (notice that the causal relations between the corresponding physical measurement actions depend on the observer performing them), then we will say that the observables $A,B$ are compatible if the outcome of $\langle B,A:S\rangle$ is $(b,a)$.
A family of observables  $\mathbf{A} = \{ A_1, \ldots, A_N\} \subset \mathcal{A}$ is said to be compatible\footnote{Notice that such family doesn't need to be finite even if, for practical purposes, we will be working under this assumption.} if they are compatible among them or, in other words, if the outcomes of their measurements do not depend on the order in which the measurements are performed.

The formulation presented in what follows is inspired by Schwinger's construction of the ``algebra of measurements'' \cite{Sc59}, where the starting point of the description of a quantum system is  the selection of a family $\mathbf{A}$ of compatible observables, e.g., the $z$-projection of the spin for a two-level system.  The outcomes of a measurement of such observables will be called in what follows just `outcomes' or `events'\footnote{Not to be confused with space-time events or with Sorkin's notion of events as subsets of the space of histories of a system \cite{So16}.   Schwinger used the term `states' for such notion, but we rather use a different terminology not to create confusion with the proper notion of states of the system as positive normalized functionals on the algebra of observables of the theory, see \cite{Ib18}.}, and we will denote them as $a,a',a'',\ldots$, etc.

At this stage, and in what follows, we will not try to make precise the meaning of `measurement' or the nature of the outcomes as we will consider them primary notions determined solely by the experimental setting used to study our system.  Neither will we require any particular algebraic structure for the family of observables used in such setting. For instance, the outcomes $a, a', a'',\ldots$,  could be just collections of real numbers.     We will just be concerned with the structural relations among the various notions that are introduced.  A concrete realisation of them will be offered in the next section by adapting Schwinger's framework to the language developed here.  

In the incipit of his first note \cite{Sc59}, Schwinger writes:

\medskip

\textit{``The classical theory of measurement is implicitly based upon the concept of an interaction between the system
of interest and the measuring apparatus that can be made arbitrarily small, or at
least precisely compensated, so that one can speak meaningfully of an idealized experiment that disturbs no property of the system. The classical representation
of physical quantities by numbers is the identification of all properties with the
results of such non-disturbing measurements. It is characteristic of atomic phenomena, however, that the interaction between system and instrument cannot be
indefinitely weakened. Nor can the disturbance produced by the interaction be
compensated precisely since it is only statistically predictable. Accordingly, a measurement on one property can produce unavoidable changes in the value previously
assigned to another property, and it is without meaning to ascribe numerical values
to all the attributes of a microscopic system. The mathematical language that is
appropriate to the atomic domain is found in the symbolic transcription of the laws
of microscopic measurement."}

\medskip 

Inspired by this remark, we postulate that for a given physical system there are transitions among the outcomes of measurements, that is, the outcome $a$ of the measurement of an observable in $\mathbf{A}$  is compatible with other different values of other observables in $\mathbf{A}$ before the act of measurement.
Such transitions are determined completely by the intrinsic dynamic of the system and by the interaction of the experimental setting with it.

Consistently with the notation introduced for groupoids, we will denote such transitions using a diagrammatic notation as $\alpha \colon a \to a'$, and we will say that the event $a$ is the source of the transition $\alpha$ and the event $a'$ is its target.  We will also say that the transition $\alpha$ transforms the outcome $a$ into the outcome $a'$. 

The allowed physical transitions $\alpha$ must satisfy a small number of natural requirements or axioms. 
The first one is that transitions can be composed, that is, if $\alpha \colon a \to a'$, and $\beta\colon a' \to  a''$ denote two allowed transitions, there is a transition $\beta \circ \alpha \colon a \to a''$.  Notice that not all transitions can be composed, we may only compose compatible transitions\footnote{In \cite{Ib18} it will be discussed the meaning of Schwinger's compound measurements, that is, the meaning of composing `incompatible' transitions.
For the purposes of this paper though, we will restrict ourselves to consider the composition of compatible transitions.}, that is, transitions $\alpha$, $\beta$ such that $\beta$ transforms the target event $a'$ of $\alpha$.   This composition law must be associative, that is, if $\alpha$, $\beta$ are transitions as before, and $\gamma \colon a'' \to a'''$, then
$$
(\gamma \circ \beta) \circ \alpha = \gamma \circ (\beta \circ \alpha ) \, .
$$
Moreover we will assume that there are trivial transitions, that is, transitions $1_a \colon a \to a$ such that $1_{a'} \circ \alpha = \alpha$ and $\alpha \circ 1_a = \alpha$ for any transition $\alpha \colon a \to a'$.  The physical meaning of transitions $1_a$ is that of manipulations of the system which do not change the outcome of any measurement of $\mathbf{A}$.  

Eventually, we will assume a (local) reversibility property of physical systems, that is, transitions are invertible: given any transition $\alpha \colon a \to a'$, there is another one $\alpha' \colon a' \to a$ such that $\alpha \circ \alpha' = 1_{a'}$ and $\alpha' \circ \alpha = 1_a$.   We will denote such transition as $\alpha^{-1}$ and it clearly must be unique\footnote{Notice that in this picture the reversibility condition could be lifted, in which case we will be dealing with categories, to include open systems.}.   

From the previous axioms, it is clear that the collection of transitions $\alpha$ and events $a$ form a groupoid as explained in Sect. \ref{sec:groupoids}.   Such groupoid will be denoted by $\mathbf{G}_{\mathbf{A}}$, its objects can be understood as the outcomes $a$ provided by measurements of a family of compatible observables $\mathbf{A}$, and its morphisms are the allowed physical transitions among events.  We will denote by $\mathbf{G}_\mathbf{A}(a,a')$ the family of transitions $\alpha \colon a \to a'$ between the events $a$ and $a'$.

We will denote by $\Omega$ the collection of all outcomes, and by $s,t$ the natural source and target maps.  Notice that so far we are not assuming that the groupoid $\mathbf{G}_{\mathbf{A}}$ has any particular additional property, that is, it could be connected or not, it could possess an Abelian isotropy group or the isotropy group could be trivial (see later on, and Sect. \ref{sec:examples} for a few concrete instances).

We will associate with the system a Hilbert space $\mathcal{H}_{A}$.   Such Hilbert space $\mathcal{H}_{A}$ is the support of the fundamental representation of the groupoid $\mathbf{G}_\mathbf{A}$.   In Sect. \ref{sec:fundamental}, the case of finite groupoids was discussed.   There, the Hilbert space was finite-dimensional and given explicitly as: 
$$
\mathcal{H}_{A} = \bigoplus_{a \in \Omega} \mathbf{C} |a \rangle .
$$
In a more general situation, we will assume that the space of events $\Omega$ is a standard Borel space with measure $\mu$.  In that case, $\mathcal{H}_{A}$ is the direct integral of the field of Hilbert spaces $\mathbb{C} |a \rangle$:
$$
\mathcal{H}_{A} = \int_{\Omega}^\oplus \mathbb{C}\,  | a \rangle \, d\mu (a) \cong L^2(\Omega, \mu) \, .
$$
For instance, in the particular case when $\Omega$ is discrete countable, such measure will be the standard counting measure and $\mathcal{H}_{A}\cong L^2(\mathbb{Z}) = l^2$.    Notice that if the space of objects is countable, the unit elements $1_a$ are represented in the fundamental representation as the orthogonal projectors $P_a$ on the subspaces $\mathbb{C} | a \rangle$. Such projectors provide a resolution of the identity for the Hilbert space $\mathcal{H}_{A}$.  

The Hilbert space $\mathcal{H}_A$ will allow us to relate the picture provided by the groupoid $\mathbf{G}_\mathbf{A}$ with the standard Dirac-Schr\"odinger picture and, in the particular instance of a discrete, finite space of events considered by Schwinger, it becomes a finite dimensional space corresponding to a finite-level quantum system.

All the previous arguments can be repeated when considering another system of observables, say $\mathbf{B}$, to describe the transitions of the system.    The system $\mathbf{B}$ may be incompatible with $\mathbf{A}$, however, they may have common events, that is events that are outcomes of both $\mathbf{A}$ and $\mathbf{B}$.   Thus, in addition to the transitions $\alpha \colon a \to_A a'$ among outcomes of the family $\mathbf{A}$, and $\beta \colon b \to_B b'$ among the outcomes of the family $\mathbf{B}$, new transitions could be added to the previous ones, i.e., those of the form $\gamma = \alpha\circ \beta \colon a \to_A a' = b \to_B b'$ where $a' = b$ is a common outcome for both $\mathbf{A}$ and $\mathbf{B}$.

We may form the groupoid $\mathbf{G}$ consisting of the groupoid generated by the union of all groupoids $\mathbf{G}_{\mathbf{A}}$ over the total space of events $\mathcal{S}$.   Each groupoid $\mathbf{G}_\mathbf{A}$ is a subgroupoid of the groupoid $\mathbf{G}$.  
The construction of $\mathbf{G}$ depends on the possibility of giving a description of a physical systems in terms of two or more maximal families of compatible observables. This instance naturally carries with it the possibility of relating different descriptions, and this, in turns, will lead us to the construction of another layer on our abstract groupoid structure. Specifically, we will build another groupoid $\boldsymbol{\Gamma}$ over $\mathbf{G}$ obtaining what is known as a 2-groupoid.
If the groupoid $\mathbf{G}$ describing the system is connected or transitive, we will say that there are no superselection rules.  The connected components of the groupoid will determine the different sectors of the theory.    In what follows, we will restrict ourselves to consider a connected component of the total groupoid. 

%%%%%%%%%%%%%%%%%%

\subsection{Two examples: the qubit and the groupoid of tangles}

\subsubsection{The qubit}

Let us start the discussion of some simple examples by considering what is arguably the simplest non-trivial groupoid structure. We call it the extended singleton or groupoid $\mathbf{A}_2$ (see \cite[Chap. 1]{Ib19} for more details), and it is given by the diagram in Fig. \ref{singleton} below:

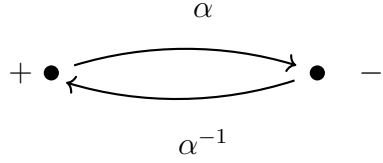
\begin{figure}[h]
\centering
\begin{tikzpicture} 
\fill (0,0) circle  (0.1);
\fill (3.5,0) circle  (0.1);
\draw [thick,->] (0.3,0.1) arc (107:73:5);
\draw [thick,->] (3.2,-0.1) arc (288:253:5);
\node [left]  at  (-0.1,0)   {$+$};
\node [right] at (3.9,0)   {$-$};
\node [above] at (2,0.6)   {$\alpha$};
\node [below] at (2,-0.6)   {$\alpha^{-1}$};
\end{tikzpicture}
\caption{The extended singleton.}
\label{singleton}
\end{figure}

This diagram will correspond to a physical system described by a family of observables $\mathbf{A}$ producing just two outputs denoted respectively by $+$ and $-$, and with just one transition $\alpha \colon + \to -$ among them.   Notice that the groupoid $\mathbf{G}_{\mathbf{A}}$ associated with this diagram has 4 elements $\{ 1_+, 1_-, \alpha, \alpha^{-1}\}$ and the space of events is just $\Omega_A = \{+,-\}$.  The corresponding (non-commutative) groupoid algebra is a complex vector space of dimension 4 generated by $e_1= 1_+$, $e_2= 1_-$, $e_3 = \alpha$ and $e_4= \alpha^{-1}$ with structure constants given by the relations:
$$
\begin{array}{llll}
e_1^2 = e_1\, , & e_2^2 = e_2 \, , & e_1 e_2 = e_2 e_1 = 0 \, , & e_3 e_4 = e_2\, , \\ 
e_4 e_3 = e_1\, , & e_3 e_3  = e_4 e_4 = 0\, , & e_1e_3 = 0\, , & e_3e_1 = e_1\, , \\
e_4 e_1 = 0\, ,  &  e_1 e_4 = e_4  \, , &  e_3e_2 = 0 \, , &  e_2e_3 = e_3 \, .
\end{array}
$$
The fundamental representation of the groupoid algebra is supported on the 2-dimensional complex space $\mathcal{H} = \mathbb{C}^2$ with canonical basis $|+\rangle$, $|-\rangle$.  The groupoid elements are represented by operators acting on the canonical basis as:
$$
A_+ |+\rangle = \pi (1_+) |+\rangle = |+\rangle \, , \qquad  A_+|-\rangle  =  \pi(1_+) |- \rangle = 0 \, ,
$$
etc., that is, for instance, the operator $A_+$ has associated matrix:
$$
A_+ = \left[ \begin{array}{cc} 1 & 0 \\ 0 & 0\end{array}\right] \, .
$$
Similarly we get:
$$
A_- = \pi(1_-) = \left[ \begin{array}{cc} 0 & 0 \\ 0 & 1\end{array}\right] \, , \quad A_\alpha =  \pi(\alpha) = \left[ \begin{array}{cc} 0 & 0 \\ 1 & 0\end{array}\right] \, , \quad A_{\alpha^{-1}} = \pi(\alpha^{-1}) =  \left[ \begin{array}{cc} 0 & 1 \\ 0 & 0 \end{array}\right] \, , 
$$
Therefore, the groupoid algebra can be naturally identified with the algebra of $2\times 2$ complex matrices $M_2(\mathbb{C})$ and the fundamental representation is just provided by the matrix-vector product of matrices and 2-component column vectors of $\mathbb{C}^2$.    The dynamical aspects of this system will be described extensively in \cite{Ib18}. 

Before discussing the next example it is interesting to observe that if we consider the system without the transition $\alpha$, that is, now the groupoid will consists solely of the elements $\{ 1_+, 1_-\}$, its corresponding groupoid algebra will be the just the 2-dimensional Abelian algebra defined by the relations: $e_1^2 = e_1$,  $e_2^2 = e_2$ and $e_1e_2 =e_2 e_1 = 0$, that is, the classical bit.  

According to the physical interpretation of transitions given at the beginning of this section, by disregarding the transitions $\alpha$ and $\alpha^{-1}$ we are implicitely assuming that experimental devices do not influence the system, which is an assumption of genuinely classical flavour.

\subsubsection{The homotopy groupoid}
An interesting family of groupoids have its origin in topology.    Consider a closed (compact without boundary), connected smooth manifold $X$ and the groupoid $\mathbf{G}(X)$ of unparametrised, oriented, piecewise smooth maps $\gamma \colon [0,1] \to X$, that will be called oriented histories on $X$.  In other words, a morphism $[\gamma]$ in $\mathbf{G}(X)$ is an equivalence class of piecewise smooth maps up to reparametrizations by positive changes of parameter $dt/ds > 0$.   The source and target maps $s,t \colon \mathbf{G}(X) \rightrightarrows X$ are defined as $s([\gamma]) = \gamma(0)$ and $t([\gamma]) = \gamma(1)$.  In what follows, we will just denote by $\gamma$ the equivalence class $[\gamma]$.   The unit morphisms are defined by the curves $1_x (t) = x$ for all $t\in [0,1]$.   The composition law is given by the standard composition of paths, that is $\gamma_1 \circ \gamma_2 (t) = \gamma_1(2t)$, $t\in [0,1/2)$ and $\gamma_1 \circ \gamma_2 (t) = \gamma_2(2t-1)$, $t\in [1/2,1]$.   

The groupoid of oriented histories $\mathbf{G}(X)$, used in \cite{Za83} to provide a universal setting to describe Lagrangian systems with topological obstructions, happens to be too large for the purposes of Topology and is drastically reduced by introducing an equivalence relation on it.   A transformation $\varphi \colon \gamma \Rightarrow \gamma'$ between two oriented histories is provided by ambient isotopies, that is,  maps $\varphi \colon [0,1]\times M \to X$  such that $\varphi(0,\gamma(t)) = \gamma(t)$, $\varphi(1,\gamma(t)) = \gamma'(t)$, $\varphi(s,t )$ is smooth in the variable $s$ and piecewise smooth on the variable $t$, and $\varphi(s,\cdot)$ is a diffeomorphism for every $s\in [0,1]$.    

The notion of the transformation $\varphi$ between groupoid elements will be used in an abstract setting in the following.    In the context of the present example, it suffices to use it to introduce an equivalence relation in the groupoid of oriented histories, whose quotient space is a groupoid over $X$, called the homotopy or Poincar\'e groupoid, whose isotropy group at $x\in X$ are isomorphic to Poincar\'e's homotopy groups $\pi_1(X,x)$ (notice that the Poincar\'e groupoid is connected iff the space $X$ is connected). 

Both the groupoid algebra and the fundamental representation of the groupoid of oriented histories $\mathbf{G}(X)$ are hard to describe.     The groupoid algebra can be described as the completion with respect to an appropriate norm of the algebra of finite formal linear combinations of oriented histories $\boldsymbol{\gamma} = \sum c_\gamma \gamma$.   
Assuming that $X$ is orientable choosing an auxiliary volume form on $X$, we may construct a measure on it and define the Hilbert space $\mathcal{H}$ of square integrable functions on $X$.    The fundamental representation of this algebra will be suported on the space of distributions on $X$ by means of $\pi(\gamma) (\delta_x) = \delta_y$ if the oriented history $\gamma$ takes $x$ into $y$, where $\delta_x$ is the Dirac's delta distribution at $x$.

More interesting is the natural generalization of the groupoid $\mathbf{G}(X)$ provided by the groupoid of braids on $X\times [0,1]$, that is, the quotient with respect to the same generalized homotopy equivalence relation of the space of $n$ non-intersecting oriented histories with end points on the boundary $\partial (X\times [0,1]) = X \times \{ 0,1\}$.    The corresponding quotient space $\mathbf{B}_n(X)$ with respect to ambient isotopies is again a groupoid over $X^n$. The fundamental representation of this groupoid is supported in the Hilbert space $L^2(X)^{\otimes n}$, and it provides relevant information on the statistics of the system described by it.  These and other aspects of the use of groupoids in the description of variational principles in quantum mechanics will be described elsewhere.

%%%%%%%%%%%%%%%%%%
%%%%%%%%%%%%%%%%%%

\subsection{The 2-groupoid structure: transformations}\label{sec:transformations}

The dynamical behaviour of a system is described by a sequence of transitions $w = \alpha_1\alpha_2\cdots \alpha_r$ that will be called histories.  In a certain limit\footnote{Notice that the limit cannot be obtained by just physically obtaining the outcomes of complete measurements because of Zeno's effect (see for instance \cite{Fa00} and references therein).}, any history would define a one-parameter family $\alpha_t$ of transitions (but we may very well keep working with discrete sequences).   

Typically, once the family of observables has been equipped with an algebra structure, these sequences of transitions are generated by a given observable  promoted to be the infinitesimal generator of a family of automorphisms.   
 However, we will not enter here in the discussion of this dynamical notions that will be the main subject of \cite{Ib18}.  What we would like to stress here is that the explicit expression of such sequence of transitions depends on the complete measurements chosen to describe the behaviour of the system and it may look very different when observed using two different systems of observables $\mathbf{A}$ and $\mathbf{B}$.   

The existence of such alternative descriptions imply the existence of families of `transformations' among transitions that would allow to compare the descriptions of the dynamical behaviour of the system (and its kinematical structure as well) when using different measurements systems.   We will also use a diagrammatic notation to denote transformations: $\varphi \colon \alpha \Rightarrow \beta$, or graphicallys as in Fig. \ref{transformations} below.

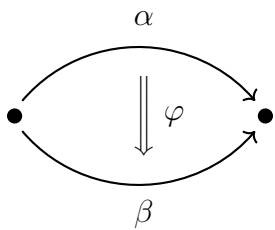
\begin{figure}[ht]
\centering
\begin{tikzpicture} 
\fill (0.2,0) circle  (0.1);
\fill (3.5,0) circle  (0.1);
\draw [thick,->] (0.3,0.2) arc (140:40:2);
\draw [thick,->] (0.3,-0.2)  arc (40:140:-2);
%\node [left]  at  (-0.1,0)   {$v_1$};
%\node [right] at (4.4,0)   {$v_2$};
\node  at (1.9,1.3)   {$\alpha$};
\node at (1.9,-1.3)   {$\beta$};
\node  at (1.9,0) {$\bigg\Downarrow$};
\node  at (2.3,0) {$\varphi$};
\end{tikzpicture}
\caption{A transformation $\varphi$ between the transitions $\alpha$ and $\beta$.}
\label{transformations}
\end{figure}

The transformations $\varphi$ must satisfy some obvious axioms.  First, in order to make sense of the assignment $\alpha  \Rightarrow \beta$, the transformation $\varphi$ must be compatible with the source and target maps of the groupoid or, in other words, $\varphi$ must map $\mathbf{G}_\mathbf{A}(a,a')$ to $\mathbf{G}_\mathbf{B}(b,b')$.  Moreover, 
the transformations $\varphi\colon \alpha \Rightarrow  \beta$ and $\psi\colon \beta \Rightarrow \gamma$ could be composed providing a new transformation $\varphi\circ_v \psi \colon \alpha \Rightarrow \gamma$, from transition $\alpha$ to $\gamma$.   This composition law will be called the `vertical' composition law and denoted accordingly by $\circ_v$ (see Fig. \ref{vertical_horizontal}(a)  for a diagrammatic representation of the vertical composition of transformations). This vertical composition law of transformations must be associative as there is no a preferred role for the various arrangements of compositions between the various transformations involved. That is, we postulate:
$$
\varphi \circ_v (\psi \circ_v \zeta) = (\varphi \circ_v \psi) \circ_v \zeta \, ,
$$
for any three transformations $\varphi \colon \alpha \Rightarrow \beta$, $\psi \colon \beta \Rightarrow \gamma$, $\zeta \colon \gamma \Rightarrow \delta$.

The transformation $\varphi$ sends the identity transition $1_a$ into the identity transition $1_b$, and there should be transformations $1_\alpha$, $1_\beta$ such that $1_\alpha \circ_v \varphi = \varphi$ and $\varphi \circ_v 1_\beta = \varphi$.   
  
Moreover, it will be assumed that transformations $\varphi \colon \alpha \Rightarrow \beta$ are reversible provided that the physical information determined by using the family of observables $\mathbf{A}$, that is, the groupoid $\mathbf{G}_\mathbf{A}$, is equivalent to that provided by the groupoid $\mathbf{G}_\mathbf{B}$, i.e., by the family of observables $\mathbf{B}$.  In other words, the transformations $\varphi \colon \alpha \Rightarrow \beta$ are invertible because there is no  natural preference among the corresponding measurements systems\footnote{In other words, we are selecting a class of equivalent measurement systems, leaving open the possibility of introducing other classes of measurements subordinate to them. These issues will be discussed in \cite{Ci19}.}.  Thus, under such assumption, for any transformation $\varphi \colon \alpha \Rightarrow \beta$ there exists another one $\varphi^{-1} \colon \beta \Rightarrow \alpha$ such that $\varphi\circ_v \varphi^{-1} = 1_\alpha$ and $\varphi^{-1}\circ_v \varphi = 1_\beta$.

Furthermore, notice that if we have a transformation $\varphi \colon \alpha \Rightarrow \beta$ and another one $\varphi' \colon \alpha' \Rightarrow \beta'$ such that $\alpha$ and $\alpha'$ can be composed, then $\beta$ and $\beta'$ will be composable too because of the consistency condition for transformations, that is, given two transitions $\alpha$, $\alpha'$ that can be composed, if the reference description for them is transformed, then the corresponding description of the transitions $\beta$ and $\beta'$  will be composable too and their composition must be the composition of the transformation of the original transitions (see Fig. \ref{vertical_horizontal}(b) for a diagrammatic description).  In other words, there will be a natural transformation between the transition $\alpha \circ \alpha'$ to the transition $\beta\circ \beta'$ that will be denoted by $\varphi \circ_h \varphi'$ and called the `horizontal' composition.   Figure \ref{vertical_horizontal} provides a diagrammatic representation of both operations $\circ_v$ and $\circ_h$:

\begin{figure}[ht]

\begin{tikzpicture} 
\fill (0,0) circle  (0.1);
\fill (3.3,0) circle  (0.1);
\draw [thick,->] (0.1,0.2) arc (140:40:2);
\draw [thick,->] (0.1,-0.2)  arc (40:140:-2);
%\node [left]  at  (-0.1,0)   {$v_1$};
%\node [right] at (4.4,0)   {$v_2$};
\node  at (1.7,1.3)   {$\alpha$};
\node at (1.7,-1.3)   {$\gamma$};
\node  at (1.7,0) {$\bigg\Downarrow$};
\node  at (2.4,0) {$\varphi\circ_v \psi$};

\node  at (3.8,0) {$\cong$};
\node  at (3.9,-2.7) {(a) Diagrammatic representation of the vertical composition law $\circ_v$.};

\fill (4.2,0) circle  (0.1);
\fill (7.5,0) circle  (0.1);
\draw [thick,->] (4.3,0.2) arc (160:20:1.7);
\draw [thick,->] (4.3,-0.2)  arc (20:160:-1.7);
\draw [thick,->] (4.4,0) -- (7.2,0);

\node  at (5.9,1.6)   {$\alpha$};
\node at (5.9,-1.6)   {$\gamma$};
\node  at (5.9,0.7) {$\Big\Downarrow$};
\node  at (6.3,-0.7)   {$\beta$};
\node  at (5.9,-0.7) {$\Big\Downarrow$};
\node  at (6.3,0.7) {$\varphi$};
\end{tikzpicture}

%\newline

\begin{tikzpicture} 
\fill (-1.4,0) circle  (0.1);
\fill (1.9,0) circle  (0.1);
\draw [thick,->] (-1.3,0.2) arc (140:40:2);
\draw [thick,->] (-1.3,-0.2)  arc (40:140:-2);
%\node [left]  at  (-0.1,0)   {$v_1$};
%\node [right] at (4.4,0)   {$v_2$};
\node at (0.4,1.3)   {$\alpha\circ \alpha'$};
\node at (0.4,-1.3)   {$\beta\circ\beta'$};
\node  at (0,0) {$\bigg\Downarrow$};
\node  at (1,0) {$\varphi\circ_h\varphi'$};

\node  at (2.6,0) {$\cong$};
\node  at (4.1,-2.7) {(b) Diagrammatic representation of the horizontal composition law $\circ_h$.};

\fill (3.2,0) circle  (0.1);
\fill (6.5,0) circle  (0.1);
\draw [thick,->] (3.3,0.2) arc (140:40:2);
\draw [thick,->] (3.3,-0.2)  arc (40:140:-2);
%\node [left]  at  (-0.1,0)   {$v_1$};
%\node [right] at (4.4,0)   {$v_2$};
\node  at (4.9,1.3)   {$\alpha$};
\node at (4.9,-1.3)   {$\beta$};
\node  at (4.9,0) {$\bigg\Downarrow$};
\node  at (5.3,0) {$\varphi$};

\fill (6.5,0) circle  (0.1);
\fill (9.8,0) circle  (0.1);
\draw [thick,->] (6.6,0.2) arc (140:40:2);
\draw [thick,->] (6.6,-0.2)  arc (40:140:-2);
%\node [left]  at  (-0.1,0)   {$v_1$};
%\node [right] at (4.4,0)   {$v_2$};
\node  at (8.2,1.3)   {$\alpha'$};
\node at (8.2,-1.3)   {$\beta'$};
\node  at (8.2,0) {$\bigg\Downarrow$};
\node  at (8.6,0) {$\varphi'$};

\end{tikzpicture}

\caption{A diagrammatic representation of vertical and horizontal composition of transformations.}
\label{vertical_horizontal}
\end{figure}

It is clear that the horizontal composition law is associative too, i.e., 
$$
\varphi \circ_h (\varphi' \circ_h \varphi'') = (\varphi \circ_h \varphi') \circ_h \varphi'' \, ,
$$
for any three transformations $\varphi \colon \alpha \Rightarrow \beta$, $\varphi' \colon \alpha' \Rightarrow \beta'$, $\varphi'' \colon \alpha'' \Rightarrow \beta''$ such that $\alpha \colon a \to a'$, $\beta \colon b \to b'$, $\alpha' \colon a' \to a''$, $\beta'\colon b' \to b''$ and $\alpha'' \colon a'' \to a'''$, $\beta'' \colon b'' \to b'''$.   The horizontal composition rule has natural units, that is, if $\varphi \colon \alpha \Rightarrow \beta$, then the transformation $1_{a'b'} \colon 1_{a'} \Rightarrow 1_{b'}$, if $\alpha \colon a \to a'$ and $\beta \colon b \to b'$, is such that: $\varphi \circ_h 1_{a'b'} = \varphi$ and $1_{ab} \circ_h \varphi = \varphi$.   

\begin{figure}[ht]
\centering
\begin{tikzpicture} 
%first blow
\fill (0.2,0) circle  (0.1);
\fill (3.5,0) circle  (0.1);
\draw [thick,->] (0.3,0.2) arc (160:20:1.7);
\draw [thick,->] (0.3,-0.2)  arc (20:160:-1.7);
\draw [thick,->] (0.4,0) -- (3.2,0);
\node  at (1.9,1.6)   {$\alpha$};
\node at (1.9,-1.6)   {$\gamma$};
\node  at (1.9,0.7) {$\Big\Downarrow$};
\node  at (2.3,-0.7)   {$\psi$};
\node  at (1.9,-0.7) {$\Big\Downarrow$};
\node  at (2.3,0.7) {$\varphi$};
\node  at (3,-2) {$\cong$};

%\node  at (3.9,0) {$\cong$};
% second blow
\fill (3.5,0) circle  (0.1);
\fill (6.8,0) circle  (0.1);
\draw [thick,->] (3.6,0.2) arc (160:20:1.7);
\draw [thick,->] (3.6,-0.2)  arc (20:160:-1.7);
\draw [thick,->] (3.7,0) -- (6.5,0);
\node  at (5.5,1.6)   {$\alpha'$};
\node at (5.2,-1.6)   {$\gamma'$};
\node  at (5.2,0.7) {$\Big\Downarrow$};
\node  at (5.6,-0.7)   {$\psi'$};
\node  at (5.2,-0.7) {$\Big\Downarrow$};
\node  at (5.6,0.7) {$\varphi'$};

\node  at (7.15,0) {$\cong$};

%first blow after composition
\fill (7.5,0) circle  (0.1);
\fill (10.8,0) circle  (0.1);
\draw [thick,->] (7.6,0.2) arc (140:40:2);
\draw [thick,->] (7.6,-0.2)  arc (40:140:-2);
%\node [left]  at  (-0.1,0)   {$v_1$};
%\node [right] at (4.4,0)   {$v_2$};
\node  at (9.2,1.3)   {$\alpha$};
\node at (9.2,-1.3)   {$\gamma$};
\node  at (9.2,0) {$\bigg\Downarrow$};
\node  at (9.9,0) {$\varphi\circ_v \psi$};
\node  at (9.8,-2) {$\cong$};

%second blow after composition
\fill (10.8,0) circle  (0.1);
\fill (14.1,0) circle  (0.1);
\draw [thick,->] (10.9,0.2) arc (140:40:2);
\draw [thick,->] (10.9,-0.2)  arc (40:140:-2);
%\node [left]  at  (-0.1,0)   {$v_1$};
%\node [right] at (4.4,0)   {$v_2$};
\node  at (12.5,1.3)   {$\alpha'$};
\node at (12.5,-1.3)   {$\gamma'$};
\node  at (12.5,0) {$\bigg\Downarrow$};
\node  at (13.3,0) {$\varphi'\circ_v \psi'$};
\end{tikzpicture}

%\newline

%first blow second line
\begin{tikzpicture} 
\fill (-0.3,0) circle  (0.1);
\fill (3,0) circle  (0.1);
\draw [thick,->] (-0.2,0.2) arc (160:20:1.7);
\draw [thick,->] (-0.2,-0.2)  arc (20:160:-1.7);
\draw [thick,->] (-0.1,0) -- (2.7,0);
\node  at (1.4,1.6)   {$\alpha\circ \alpha'$};
\node at (1.4,-1.6)   {$\gamma\circ \gamma'$};
\node  at (1.2,0.6) {$\Big\Downarrow$};
\node  at (2,-0.5)   {$\psi\circ_h \psi'$};
\node  at (1.2,-0.6) {$\Big\Downarrow$};
\node  at (2.05,0.5) {$\varphi\circ_h \varphi'$};

\node  at (3.9,0) {$\cong$};

%final blow second line
\fill (4.7,0) circle  (0.1);
\fill (8,0) circle  (0.1);
\draw [thick,->] (4.8,0.2) arc (140:40:2);
\draw [thick,->] (4.8,-0.2)  arc (40:140:-2);
%\node [left]  at  (-0.1,0)   {$v_1$};
%\node [right] at (4.4,0)   {$v_2$};
\node  at (6.4,1.3)   {$\alpha\circ \alpha'$};
\node at (6.4,-1.3)   {$\beta\circ\beta'$};
\node  at (6.4,0) {$\bigg\Downarrow$};
\node  at (10.1,-0.5) {$(\varphi\circ_h\varphi') \circ_v (\psi\circ_h\psi')$};
\node  at (10.1,0.5) {$(\varphi\circ_v\psi) \circ_h (\varphi'\circ_v\psi')$};
\node  at (10.1,0) {$=$};

\end{tikzpicture}

\caption{A diagrammatic representation of the exchange identity:  starting with the first diagram move either first row right and down or down and second row right.}
\label{exchange}
\end{figure}
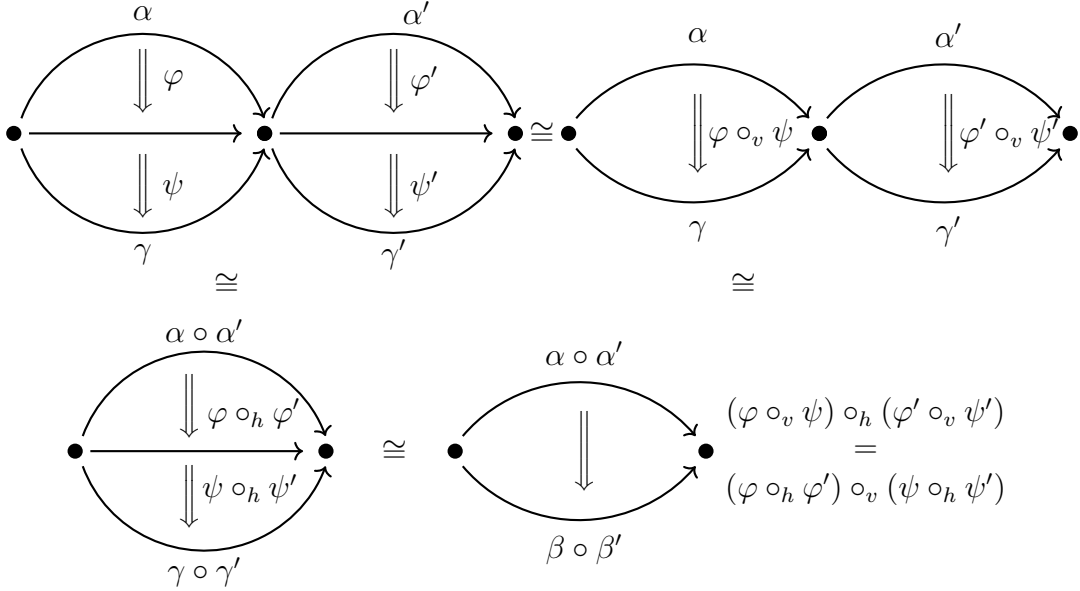

We observe that there must be a natural compatibility condition between the composition rules $\circ_v$ and $\circ_h$.   If we have two pairs of vertically composable transformations $\varphi, \psi$ and $\varphi', \psi'$ that can be also pairwise composed horizontally, then  the horizontal composition of the previously vertically composed pairs must be the same as the vertical composition of the previously horizontally composed pairs.  This consistency condition will be called the exchange identity.   Formally, it is written as follows and a diagramatic description is provided in Fig. \ref{exchange}:
\begin{equation}\label{exchange_identity}
(\varphi\circ_v \psi) \circ_h (\varphi'\circ_v \psi')  = (\varphi \circ_h \varphi') \circ_v (\psi \circ_h \psi') \, .
\end{equation}

Notice that given a transformation $\varphi \colon \alpha \Rightarrow \beta$, we can define two maps, similar to the source and target defined previously for transitions, that assign  $\alpha$ and $\beta$ to $\varphi$, respectively.    The family of invertible transformations $\boldsymbol{\Gamma}$ with the vertical composition law and the source and target maps defined in this way form again a groupoid over the space of transitions $\mathbf{G}$.  
Moreover, the source and target maps are morphisms of groupoids.  

This `double' groupoid structure, that is, a groupoid (the family of invertible transformations) whose objects (the family of all transitions) form again a groupoid (whose objects are the events) and such that the source and target maps are groupoid homomorphisms, is called a 2-groupoid.   

The morphisms of the first groupoid structure  $\boldsymbol{\Gamma}$ (or external groupoid structure) are sometimes called 2-morphisms (or 2-cells).   In our case, 2-morphisms correspond to  what we have called transformations.   The objects $\mathbf{G}$ of the first groupoid structure which are the morphisms of the second groupoid structure (or inner groupoid structure) are called 1-morphisms (or 1-cells).  In our setting, they will correspond to what we have called transitions.  Eventually, the objects $S$ of the second groupoid structure are called 0-morphisms (or 0-cells) and in the discussion before they correspond to  what we have called events.
 
The set of axioms discussed above can be thus summarised by saying that the notions previously introduced to describe a physical system form a 2-groupoid with 0-, 1- and 2-cells being, respectively, events, transitions and transformations. Therefore, if we denote the total  2-groupoid by $\boldsymbol{\Gamma}$, we will denote its set of 1-morphisms by $\mathbf{G}$, and the outer groupoid structure will be provided by the source and target maps $\boldsymbol{\Gamma} \rightrightarrows \mathbf{G}$.  The groupoid composition law in $\boldsymbol{\Gamma}$ will be called the vertical composition law and denoted $\circ_v$.  Because $\mathbf{G} \rightrightarrows S$ is itself a groupoid and the source and target maps of $\boldsymbol{\Gamma}$ are themselves groupoid homomorphisms, we can always define a horizontal composition law in a natural way denoted by $\circ_h$ and both composition laws must satisfy the exchange identity above, Eq. (\ref{exchange_identity}).

Consequently, we conclude by postulating that the categorical description of a physical system is provided by a 2-groupoid $\boldsymbol{\Gamma}\rightrightarrows \mathbf{G} \rightrightarrows S$. The 1-cells of the 2-groupoid will be interpreted as the transitions of the system and its 0-cells will be considered as the events or outcomes of measurements performed on the system.  The 2-cells will be interpreted as the transformations among transitions of theory providing the basis for its dynamical interpretation. 

The identification of these abstracts notions with the corresponding standard physical notions and their relations with other physical notions like quantum states, unitary transformations, etc., will be provided by constructing representations of the given 2-groupoid.   

Before elaborating on this, we will briefly discuss a particular instance where all these abstract structures have a specific physical interpretation, albeit not the only possible one, and it corresponds to that proposed by Schwinger in his original presentation of the `algebra of measurements'.

%%%%%%%%%%%%%%%%%%%%%

\section{Schwinger's algebra of selective measurements}

\subsection{Schwinger's algebra of measurements symbols}\label{sec:schwinger_algebra}
 
As discussed in the introduction in \cite{Sc59} (see \cite{Sc70} for a complete presentation) J. Schwinger described the fundamental algebraic relations satisfied by the set of symbols representing fundamental measurement processes of given quantum systems.  In this section, we will review such algebraic relations under similar simplifying assumptions used by Schwinger himself from the groupoid perspective discussed in  Sect. \ref{sec:transitions}.

Then following Schwinger and according with the discussion in the beginning of Sect. \ref{sec:transitions}, we define a selective measurement $M(\mathbf{a})$ associated with the family of compatible observables $\mathbf{A} = \{ A_1, \ldots, A_N\}$, as a process (or device) that rejects all elements $S$ of the ensemble $\mathcal{E}$ whose outcomes $\mathbf{a}' = (a'_1, \ldots, a'_N)$  are different from $\mathbf{a} = (a_1, \ldots, a_N)$ and leaving only the elements $S$  such that  the measurement of $\mathbf{A}$ on it gives $\mathbf{a}$.    Denoting the outcomes of the measurement of $\mathbf{A}$ on the element $S$ by $\langle \mathbf{A}:S\rangle$, then we may write $M(\mathbf{a})S = S$ if $\mathbf{a} = \langle \mathbf{A}:S\rangle$ and $M(\mathbf{a})S =  \emptyset$ if $\mathbf{a} \neq \langle \mathbf{A}:S\rangle$.   Hence, using a set-theoretical notation, we may define the subensemble:
$$
\mathcal{E}_{\mathbf{A}}(\mathbf{a}) = \{ S\in \mathcal{E} \mid \langle \mathbf{A}:S\rangle  = \mathbf{a} \} \, ,
$$ 
(in what follows we will just write $\mathcal{E}(\mathbf{a})$ unless risk of confusion).  Then, we may rewrite the previous definition as $M(\mathbf{a}) S = S$ if $S \in \mathcal{E}(\mathbf{a})$ and $M(\mathbf{a}) S = \emptyset$ otherwise.  The empty set $\emptyset$ will be formally added to the ensemble $\mathcal{E}$ with the physical meaning of the absence of the physical system.  Notice, however, that if the ensemble $\mathcal{E}$ is sufficient for $\mathbf{A}$ it cannot happen that $M(\mathbf{a}) S =\emptyset$ for all $S$.  Then, with this notation, we have $M(\mathbf{a}) \mathcal{E} = \mathcal{E}(\mathbf{a})$.

In what follows we will assume that the selective measurements $M(\mathbf{a})$ defined by  families $\mathbf{A}$ of compatible observables are unique.    In other words, suppose that we may prepare a different physical implementation $M'(\mathbf{a})$ of the selective measurement associated with $\mathbf{A}$  that  selects the subensemble $\mathcal{E}(\mathbf{a})$ by using different apparatuses, but then $M'(\mathbf{a})$ and $M(\mathbf{a})$ will be identified and denoted collectively as $M(\mathbf{a})$.     

Schwinger assumes that the `state' of a quantum system is established by performing on it a maximal selective measurement, and thus the space of states $\mathcal{S}$ of the theory is provided by the space of outcomes of complete selective measurements. According to the general picture introduced in the previous section, we identify Schwinger's space of states (the space of outcomes of a complete selective measurement) with the space of events\footnote{We prefer the word `event' (or just `outcome') rather than the word `state' in view of the discussion of states as positive normalized functionals on the algebra of observables of the theory, see \cite{Ib18}.}.  It is in this sense that Schwinger's picture produces a concrete realisation of the abstract description of physical systems used in the previous section, Sect. \ref{sec:systems}.
 
Again following Schwinger, a more general class of symbols $M(\mathbf{a}',\mathbf{a})$ are introduced.   They denote selective measurements that reject all elements of an ensemble whose outcomes are different from $	\mathbf{a}$ and those accepted are changed in such a way that their outcomes are $\mathbf{a}'$.  Using the set-theoretical notation introduced above we have:  $M(\mathbf{a}',\mathbf{a})S  \in \mathcal{E}(\mathbf{a}')$ if $S\in \mathcal{E}(\mathbf{a})$ and $M(\mathbf{a}',\mathbf{a})S = \emptyset$ otherwise.    Notice that consistently $M(\mathbf{a},\mathbf{a}) = M(\mathbf{a})$ and  $M(\mathbf{a}',\mathbf{a})  \mathcal{E} = \mathcal{E}(\mathbf{a}')$.

%Again as in the case of $M(\mathbf{a})$, there could be different physical devices used to implement the selective measurements $M(\mathbf{a}',\mathbf{a})$, that is, there could exists many different physical implementations $M'(\mathbf{a}',\mathbf{a})$, $M''(\mathbf{a}',\mathbf{a})$, ... ,  of the complete selective measurement that change the elements $S \in  \mathcal{E}(\mathbf{a})$ into elements $S' \in \mathcal{E}(\mathbf{a}')$ described above.   Even if it could be convenient to consider them as  different selective measurements depending on the physical processes involved in their actual implementation we will proceed by introducing the simplifiying assumption of considering them equivalent, that is we will denote simply by $M(\mathbf{a}',\mathbf{a}) $ the equivalence class of generalized selective measurements described above exclusively by the transformation taken on elements on the ensemble $\mathcal{E}$.

It is an immediate consequence of the basic definitions above that if we consider the natural composition law of selective measurements $M(\mathbf{a}'',\mathbf{a}') \circ M(\mathbf{a}',\mathbf{a})$ defined as the selective measurement obtained by performing first the selective measurement $ M(\mathbf{a}',\mathbf{a})$ and immediately afterwards the selective measurement $M(\mathbf{a}'',\mathbf{a}')$, then we get:
 \begin{equation}\label{schwinger_groupoid_law}
 M(\mathbf{a}'',\mathbf{a}') \circ M(\mathbf{a}',\mathbf{a}) = M(\mathbf{a}'',\mathbf{a})  \, , 
 \end{equation}
and 
\begin{equation}\label{schwinger_units}
 M(\mathbf{a}')\circ M(\mathbf{a}',\mathbf{a}) = M(\mathbf{a}',\mathbf{a}) \, , \qquad  M(\mathbf{a}',\mathbf{a}) \circ  M(\mathbf{a})= M(\mathbf{a}',\mathbf{a}) \, ,
\end{equation}
It is clear that performing two selective measurements $M(\mathbf{a}',\mathbf{a})$, and  $M(\mathbf{a}''',\mathbf{a}'')$ one after the other will produce a selective measurement again only if $\mathbf{a}'' = \mathbf{a}'$, otherwise if $\mathbf{a}'' \neq \mathbf{a}'$, then  $ M(\mathbf{a}''',\mathbf{a}'') \circ M(\mathbf{a}',\mathbf{a}) S = \emptyset$ for all $S$ which, as indicated before, is not a selective measurement of the form $ M(\mathbf{a}',\mathbf{a})$. 

Notice that if we have three selective measurements $M(\mathbf{a},\mathbf{a}')$,  $M(\mathbf{a}',\mathbf{a}'')$ and $M(\mathbf{a}'',\mathbf{a}''')$ then, because of the basic definitions, the associativity of the composition law holds:
\begin{equation}\label{schwinger_associative}
M(\mathbf{a},\mathbf{a}') \circ (M(\mathbf{a}',\mathbf{a}'') \circ M(\mathbf{a}'',\mathbf{a}''')) = (M(\mathbf{a},\mathbf{a}') \circ M(\mathbf{a}',\mathbf{a}'')) \circ M(\mathbf{a}'',\mathbf{a}''') \, .
\end{equation}
Finally, it is worth to observe that given a measurement symbol $M(\mathbf{a}',\mathbf{a})$ the measurement symbol $M(\mathbf{a},\mathbf{a}')$ is such that:
\begin{equation}\label{schwinger_inverse}
M(\mathbf{a}',\mathbf{a}) \circ M(\mathbf{a},\mathbf{a}') = M(\mathbf{a}') \, ,\quad M(\mathbf{a},\mathbf{a}') \circ M(\mathbf{a}',\mathbf{a}) = M(\mathbf{a}) \, .
\end{equation}

Then, we conclude that the composition law of selective measurements satisfying Eqs. \eqref{schwinger_groupoid_law}, \eqref{schwinger_units}, \eqref{schwinger_associative}, \eqref{schwinger_inverse}, determines a groupoid law in the collection $\mathbf{G}_\mathbf{A}$ of all measurement symbols $M(\mathbf{a}',\mathbf{a})$ associated with the complete family of observables $\mathbf{A}$, whose objects (the events of the system) are the possible outcomes $\mathbf{a}$ of the observables in $\mathbf{A}$.     

In this formulation, the general selective measurements $M(\mathbf{a}', \mathbf{a})$ are the morphisms of the groupoid $\mathbf{G}_\mathbf{A}$ and correspond to the notion of `transitions' introduced in the general setting described in the previous section.   Notice that, in this context, `transitions' have not a dynamical meaning but are the consequences of deliberate manipulation of the system by the observers\footnote{Of course, the observers may decide to use the Hamitonian determining the dynamical evolution, and use selective measurements that use the dynamics itself to change the system.}.   Even if at this moment we could change our notation to $\alpha \colon \mathbf{a} \to \mathbf{a}'$ as in the previous section, we will stick with Schwinger's original notation to facilitate the comparison with the original presentation.

%%%%%%%%%%%%%%%%%%

\subsection{Stern-Gerlach measurements}

We have just seen that, in Schwinger's picture of Quantum Mechanics, with a given physical system we associate a family of groupoids $\mathbf{G}_\mathbf{A}$ for all maximal families of compatible observables $\mathbf{A}$.     As it was pointed out before, events are defined as outcomes of maximal selective measurements, thus for each maximal family of compatible observables $\mathbf{A}$ there is a family of events $\mathcal{S}_{\mathbf{A}}$.      Given another different maximal family of compatible observables $\mathbf{B}$, it would determine another family of events, denoted now by $\mathbf{b} \in \mathcal{S}_\mathbf{B}$.    It could happen that the sub-ensemble determined by the selective measurement $M_\mathbf{A}(\mathbf{a})$ would lie inside that defined by $M_\mathbf{B}(\mathbf{b})$, that is $\mathcal{E}_\mathbf{A}(\mathbf{a}) \subset \mathcal{E}_\mathbf{B}(\mathbf{b})$.  We will say that in such case $\mathbf{b}$ is subordinate to $\mathbf{a}$ and we will denote it by $\mathbf{b} \subset  \mathbf{a} $.   In such case the measurement of $\mathbf{B}$ will not modify the outcomes defined by the sub-ensemble determining the event $\mathbf{a}$.    

If it happens that $\mathbf{a}$ is subordinate to $\mathbf{b}$ and viceversa, i.e., 
$ \mathbf{b} \subset  \mathbf{a} $ and $\mathbf{a} \subset  \mathbf{b}$, we will consider that both events are the same, $ \mathbf{b} \sim \mathbf{a} $, and we will treat them as the same object.   Thus the space of events $\mathcal{S}$ of the system is the collection of equivalence classes of events with respect to equivalence relation ``$\sim$'' associated with the subordination ``$\subset$'' relation among them.    It is clear now that the relation $\subset$ induces a partial order relation on the set of events $\mathcal{S}$ and can be used to induce a partial order in the space of observables.  The implications of such structure will be discussed in \cite{Ib18}.

Notice that the family $\bigcup_{\mathbf{A}}\mathbf{G}_\mathbf{A}$ of all groupoids $\mathbf{G}_{\mathbf{A}}$ associated with all  selective measurements is a groupoid over the space of events $\mathcal{S}$.   Under such premises two selective measurements $M_\mathbf{A}(\mathbf{a}',\mathbf{a})$ and  $M_\mathbf{B}(\mathbf{b}',\mathbf{b})$ can be composed if  and only if $\mathbf{a} \sim  \mathbf{b'}$ in which case:
$M_\mathbf{A}(\mathbf{a}',\mathbf{a}) \circ M_\mathbf{B}(\mathbf{b}',\mathbf{b})$
will correspond to a physical device that will take as inputs elements in the sub-ensemble $\mathcal{E}_\mathbf{B}(\mathbf{b})$ and will return elements in the sub-ensemble $\mathcal{E}_\mathbf{A}(\mathbf{a})$.    If we denote it by $M_{\mathbf{AB}}(\mathbf{a}', \mathbf{b})$ (or simply $M(\mathbf{a}', \mathbf{b})$ for short) then, provided that $ \mathbf{a} \sim \mathbf{b'} $ we get again:
$$
M(\mathbf{a}',\mathbf{a}) \circ M (\mathbf{b}',\mathbf{b}) =  M(\mathbf{a}', \mathbf{b}) \, .
$$
The operation $M(\mathbf{a}', \mathbf{b}) $ is invertible too.  Actually the operation 
$$
M(\mathbf{b}, \mathbf{a}') =   M (\mathbf{b},\mathbf{b}') \circ M(\mathbf{a},\mathbf{a}') \, ,
$$
is such that $M(\mathbf{b}, \mathbf{a}') \circ M(\mathbf{a}', \mathbf{b}) = M (\mathbf{b},\mathbf{b}') \circ M(\mathbf{a},\mathbf{a}') \circ M(\mathbf{a}',\mathbf{a}) \circ M (\mathbf{b}',\mathbf{b}) = M(\mathbf{b}) = 1_\mathbf{b}$ and $M(\mathbf{a}', \mathbf{b}) \circ M(\mathbf{b}, \mathbf{a}')  = 1_{\mathbf{a}'}$.

Notice that this new family of operations does not correspond (in general) to a set of compatible observables $\mathbf{C}$ constructed out of $\mathbf{A}$ and $\mathbf{B}$\footnote{Until an algebra structure is introduced in space of observables in which case we can start looking for maximal Abelian subalgebras generated by them \cite{Ib18}.}, however, it is important to introduce them because they correspond to the physical operations of composition of Stern-Gerlach devices (so they could be called \textit{composite Stern-Gerlach measurements} or just $SG$-measurements for short).   These extended operations were also introduced by Schwinger as a consistency condition for the relations of the algebra of selective measurements.  However, in the reformulation of the basic notions we are presenting here, they will play a more significant role as they would uncover another layer of structure in the algebraic setting for basic measurement operations that, as it was discussed in the previous section, is that of a 2-groupoid.    

All together, if we consider the space $\mathbf{G}$ consisting of all complete selective measurements and composite Stern-Gerlarch measurements together with their natural composition law, we see that it has the structure of a groupoid whose space of objects is the space of events $\mathcal{S}$ of the system.
Thus we will denote by $\mathbf{G} \rightrightarrows \mathcal{S}$ such groupoid with the source and target maps $s,t$ given by: 
$$
s(M(\mathbf{a}',\mathbf{a}) ) = \mathbf{a} \, , \qquad t(M(\mathbf{a}',\mathbf{a}) ) = \mathbf{a}'  \, .
$$

%%%%%%%%%%%%%%%%

\subsection{The 2-groupoid structure of Schwinger's algebra of selective measurements}

It is clear that SG-measurements define transformations among selective measurements.  That is, if we consider the transition $M_\mathbf{A}(\mathbf{a}, \mathbf{a}')$ and the SG-measurements $M(\mathbf{a}', \mathbf{b}')$ and $M(\mathbf{b}, \mathbf{a})$ that transform the ensembles $\mathcal{E}_\mathbf{A}(\mathbf{a}')$ in $\mathcal{E}_\mathbf{B}(\mathbf{b}')$, and $\mathcal{E}_\mathbf{B}(\mathbf{b})$ in $\mathcal{E}_\mathbf{A}(\mathbf{a})$ respectively, then the transition $M(\mathbf{b}, \mathbf{a}) \circ M_\mathbf{A}(\mathbf{a}, \mathbf{a}') \circ M(\mathbf{a}', \mathbf{b}')$ must be the transition corresponding to the selective measurement $M_\mathbf{B}(\mathbf{b}, \mathbf{b}')$, that is:
\begin{equation}\label{schwinger_transformation}
M(\mathbf{b}, \mathbf{a}) \circ M_\mathbf{A}(\mathbf{a}, \mathbf{a}') \circ M(\mathbf{a}', \mathbf{b}') = M_\mathbf{B}(\mathbf{b}, \mathbf{b}') \, .
\end{equation}
Hence, formula (\ref{schwinger_transformation}) defines a transformation $\varphi \colon M_\mathbf{A}(\mathbf{a}, \mathbf{a}')  \Rightarrow M_\mathbf{B}(\mathbf{b}, \mathbf{b}')$ in the sense of Sect. \ref{sec:transformations}.   This transformation could be just denoted as $\varphi(\mathbf{a}, \mathbf{a}'; \mathbf{b}, \mathbf{b}')$ instead of listing the pair of SG-measurements $M(\mathbf{b}, \mathbf{a}) $  and $M(\mathbf{a}', \mathbf{b}')$ involved on its definition in order to avoid a too cumbersome notation.

It is a simple matter to check the axioms introduced in Sect. \ref{sec:transformations} for the theory of transformations.   The unit transformations are given by the pairs $M(\mathbf{a})$ and $M(\mathbf{a}')$, that is, $1_{(\mathbf{a}, \mathbf{a}')} = \varphi(\mathbf{a}, \mathbf{a}'; \mathbf{a}, \mathbf{a}')$ because $\varphi(\mathbf{a}, \mathbf{a}'; \mathbf{a}, \mathbf{a}')\circ_v \varphi(\mathbf{a}, \mathbf{a}'; \mathbf{b}, \mathbf{b}')$ would transform the transition $M_\mathbf{A}(\mathbf{a}, \mathbf{a}')$ into the transition:
\begin{eqnarray*}
M_\mathbf{A}(\mathbf{a}, \mathbf{a}') & \overset{\varphi(\mathbf{a}, \mathbf{a}'; \mathbf{a}, \mathbf{a}')}{\Rightarrow} & M(\mathbf{a}) \circ M_\mathbf{A}(\mathbf{a}, \mathbf{a}') \circ M(\mathbf{a}') \\ &\overset{\varphi(\mathbf{a}, \mathbf{a}'; \mathbf{b}, \mathbf{b}')}{\Rightarrow} & M(\mathbf{b}, \mathbf{a}) \circ M_\mathbf{A}(\mathbf{a}, \mathbf{a}') \circ M(\mathbf{a}', \mathbf{b}') =  M_\mathbf{B}(\mathbf{b}, \mathbf{b}') \, .
\end{eqnarray*}

Notice that with this notation the vertical composition law is simply written as:
$$
\varphi(\mathbf{a}, \mathbf{a}'; \mathbf{b}, \mathbf{b}') \circ_v \varphi(\mathbf{b}, \mathbf{b}'; \mathbf{c}, \mathbf{c}') =  \varphi(\mathbf{a}, \mathbf{a}'; \mathbf{c}, \mathbf{c}') \, .
$$
as it is shown by the following computation:
\begin{eqnarray*}
&& M(\mathbf{c}, \mathbf{b})  	\circ \left( M(\mathbf{b}, \mathbf{a}) \circ M_\mathbf{A}(\mathbf{a}, \mathbf{a}') \circ M(\mathbf{a}', \mathbf{b}') \right)  \circ M(\mathbf{b}', \mathbf{c}') \\ &&=\left (M(\mathbf{c}, \mathbf{b})  \circ M(\mathbf{b}, \mathbf{a}) \right) \circ M_\mathbf{A}(\mathbf{a}, \mathbf{a}') \circ \left( M(\mathbf{a}', \mathbf{b}') \circ M(\mathbf{b}', \mathbf{c}') \right)   \\ && = M(\mathbf{c}, \mathbf{a}) \circ   M_\mathbf{A}(\mathbf{a}, \mathbf{a}'') \circ M(\mathbf{a}'', \mathbf{c}'')  = M_\mathbf{B}(\mathbf{c}, \mathbf{c}'') \, ,
\end{eqnarray*}

The associativity property of the vertical composition law is easily checked in a similar way.    

Regarding the horizontal composition law, we must notice that if  
$$\varphi(\mathbf{a}, \mathbf{a}'; \mathbf{b}, \mathbf{b}') \colon M_\mathbf{A}(\mathbf{a}, \mathbf{a}')  \Rightarrow M_\mathbf{B}(\mathbf{b}, \mathbf{b}') \, ,$$
and 
$$\varphi(\mathbf{a}', \mathbf{a}''; \mathbf{b}', \mathbf{b}'') \colon M_\mathbf{A}(\mathbf{a}', \mathbf{a}'')  \Rightarrow M_\mathbf{B}(\mathbf{b}', \mathbf{b}'') \, ,$$
denote two transformations, then the composition:
\begin{eqnarray*}
&& \left( M(\mathbf{b}, \mathbf{a}) \circ M_\mathbf{A}(\mathbf{a}, \mathbf{a}') \circ M(\mathbf{a}', \mathbf{b}') \right) \circ \left( M(\mathbf{b}', \mathbf{a}') \circ M_\mathbf{A}(\mathbf{a}', \mathbf{a}'') \circ M(\mathbf{a}'', \mathbf{b}'') \right) \\ &&= M(\mathbf{b}, \mathbf{a}) \circ \left( M_\mathbf{A}(\mathbf{a}, \mathbf{a}') \circ M_\mathbf{A}(\mathbf{a}', \mathbf{a}'')\right) \circ M(\mathbf{a}'', \mathbf{b}'') \\ && = M(\mathbf{b}, \mathbf{a}) \circ   M_\mathbf{A}(\mathbf{a}, \mathbf{a}'') \circ M(\mathbf{a}'', \mathbf{b}'')  = M_\mathbf{B}(\mathbf{b}, \mathbf{b}'') \, ,
\end{eqnarray*}
(where we have used that $SG$-measurements are invertible) shows that the pair of $SG$-measurements $M(\mathbf{b}, \mathbf{a})$ and $M(\mathbf{a}'', \mathbf{b}'')$ define a transformation from $M_\mathbf{A}(\mathbf{a}, \mathbf{a}'') $ to $M_\mathbf{B}(\mathbf{b}, \mathbf{b}'') $ or, in other words:
$$
\varphi(\mathbf{a}, \mathbf{a}'; \mathbf{b}, \mathbf{b}')  \circ_h \varphi(\mathbf{a}', \mathbf{a}''; \mathbf{b}', \mathbf{b}'')  = \varphi(\mathbf{a}, \mathbf{a}''; \mathbf{b}', \mathbf{b}'') \, .
$$

Finally, a simple computation shows that the exchange identity (\ref{exchange_identity}) is satisfied. That is, we compute first:
\begin{eqnarray*}
&& \left( \varphi(\mathbf{a}, \mathbf{a}'; \mathbf{b}, \mathbf{b}')  \circ_h \varphi(\mathbf{a}', \mathbf{a}''; \mathbf{b}', \mathbf{b}'')  \right) \circ_v 
\left( \varphi(\mathbf{b}, \mathbf{b}'; \mathbf{c}, \mathbf{c}')  \circ_h \varphi(\mathbf{b}', \mathbf{b}''; \mathbf{c}', \mathbf{c}'')  \right) \\ && = 
 \varphi(\mathbf{a}, \mathbf{a}''; \mathbf{b}, \mathbf{b}'')  \circ_v
 \varphi(\mathbf{b}, \mathbf{b}''; \mathbf{c}, \mathbf{c}'') =  \varphi(\mathbf{a}, \mathbf{a}''; \mathbf{c}, \mathbf{c}'') \, , 
\end{eqnarray*}
but
\begin{eqnarray*}
&& \left( \varphi(\mathbf{a}, \mathbf{a}'; \mathbf{b}, \mathbf{b}')  \circ_v \varphi(\mathbf{b}, \mathbf{b}'; \mathbf{c}, \mathbf{c}')  \right) \circ_h 
\left( \varphi(\mathbf{a}', \mathbf{a}''; \mathbf{b}', \mathbf{b}'')  \circ_v \varphi(\mathbf{b}', \mathbf{b}''; \mathbf{c}', \mathbf{c}'')  \right) \\ && = 
 \varphi(\mathbf{a}, \mathbf{a}'; \mathbf{c}, \mathbf{c}')  \circ_h
 \varphi(\mathbf{a}', \mathbf{a}''; \mathbf{c}', \mathbf{c}'') =  \varphi(\mathbf{a}, \mathbf{a}''; \mathbf{c}, \mathbf{c}'')  \, .
\end{eqnarray*}
which proves the desired identity.

\medskip

The previous discussion can be summarised by saying that the family of Schwinger's maximal selective and Stern-Gerlach measurements have the structure of a 2-groupoid whose 2-cells (corresponding to the notion of transformations discussed in Sect. \ref{sec:transformations})  are given by pairs of SG-measurements. The 1-cells, or transitions, are given by maximal measurements, and its 0-cells, or events, are given by the outcomes of maximal selective measurements.

%%%%%%%%%%%%%%%%

\subsection{The fundamental representation of Schwinger's algebra of selective measurements and the standard pictures of QM}

The fundamental representation of a groupoid was introduced in Sect. \ref{sec:fundamental}.  We will use this representation to provide an interpretation of Schwinger's 2-groupoid in terms of the standard pictures of Quantum Mechanics.     First we will assume, as Schwinger's did, that the set of outcomes $\{ \mathbf{a} \}$ of maximal measurements is finite.  This restriction can be lifted but we will not worry much about it here.   We will select the subgroupoid $\mathbf{G}_{\mathbf{A}}$ corresponding to transitions $M(\mathbf{a},\mathbf{a}')$.   The fundamental representation of the groupoid $\mathbf{G}_{\mathbf{A}}$ will take place on the finite dimensional Hilbert space $\mathcal{H}_{A}$ generated by the set of outcomes $\mathbf{a}$.    Thinking now of the events $\mathbf{a}$ as labels ordered from 1 to $n$, with $n$ the number of outcomes (and the dimension of $\mathcal{H}_{A}$), the elements in the groupoid algebra $\mathbb{C}[\mathbf{G}_{\mathbf{A}}]$ can be written as:
$$
A = \sum_{\mathbf{a}, \mathbf{a}' = 1}^n A_{\mathbf{a}, \mathbf{a}'}  M(\mathbf{a}, \mathbf{a}') \, ,
$$
i.e., they are formal linear combinations of the selective measurements with complex coefficients $A_{\mathbf{a}, \mathbf{a}'}$.   Hence, they can be identified with $n\times n$ matrices whose entries are given by $A_{\mathbf{a}, \mathbf{a}'}$.  The groupoid algebra composition law is just given by multiplication of matrices, that is:
\begin{eqnarray*}
A\cdot B &=& \sum_{\mathbf{a}, \mathbf{a}' , \mathbf{a}'', \mathbf{a}''' = 1}^n A_{\mathbf{a}, \mathbf{a}'}  B_{\mathbf{a}'', \mathbf{a}'''} \, \,    \delta (\mathbf{a}', \mathbf{a}'')\,  M(\mathbf{a}, \mathbf{a}') \circ M(\mathbf{a}'', \mathbf{a}''')  \\ &=&  \sum_{\mathbf{a},\mathbf{a}''' = 1}^n \left( \sum_{\mathbf{a}'' = 1}^n A_{\mathbf{a}, \mathbf{a}''}  B_{\mathbf{a}'', \mathbf{a}'''} \right)  M(\mathbf{a}, \mathbf{a}''') \\ &=&  \sum_{\mathbf{a},\mathbf{a}''' = 1}^n (AB)_{\mathbf{a}, \mathbf{a}'''}  M(\mathbf{a}, \mathbf{a}''')  \, ,
\end{eqnarray*}
where in the last row, $AB$ stands for the standard matrix product of the matrices $A$ and $B$.

Continuing with this interpretation, we notice that, using the canonical orthonormal basis provided by the vectors $|\mathbf{a}\rangle$, the vectors $| \psi \rangle$ in the fundamental Hilbert space $\mathcal{H}_{A}$ can be identified with column vectors with components $\psi_{\mathbf{a}}$, that is, $|\psi \rangle = \sum_{\mathbf{a} = 1}^n \psi_{\mathbf{a}} |\mathbf{a}\rangle$.  Then, we get:
$$
\pi (A) |\psi \rangle =  \sum_{\mathbf{a}, \mathbf{a}', \mathbf{a}'' = 1}^n  A_{\mathbf{a}, \mathbf{a}'}  \psi_{\mathbf{a}''}\, \, \delta (\mathbf{a}, \mathbf{a}'')\,  \pi (M(\mathbf{a}, \mathbf{a}')) |\mathbf{a}'' \rangle  = 
$$
$$
=\sum_{\mathbf{a}, \mathbf{a}' = 1}^n  A_{\mathbf{a}, \mathbf{a}'}  \psi_{\mathbf{a}}\,  |\mathbf{a}' \rangle = \sum_{ \mathbf{a}' = 1}^n  (A \psi)_{\mathbf{a}'} \,  |\mathbf{a}' \rangle \, ,
$$
where $A\psi$ in the last row denotes the standard matrix-vector product.  Hence, the fundamental  representation becomes just the standar representation of the algebra of $n\times n$ matrices on the corresponding linear space $\mathbb{C}^n$.   

Vectors in the fundamental space can be identified with complex linear combinations of events and transitions are represented by rank 1 operators.  In other words the transition $M(\mathbf{a}, \mathbf{a}')$ is represented by the operator $|\mathbf{a}' \rangle \langle \mathbf{a} |$ in $\mathcal{H}_{A}$.  Elements $A$ in the groupoid algebra are then represented as operators acting on $\mathcal{H}_{A}$.  
In this sense the units $1_\mathbf{a}$ of the groupoid $\mathbf{G}_{\mathbf{A}}$ are represented by the rank one orthogonal projectors $|\mathbf{a} \rangle \langle \mathbf{a} |$ that provide a resolution of the identity of $\mathcal{H}_{A}$, $ \sum_{\mathbf{a} = 1}^n | \mathbf{a} \rangle \langle \mathbf{a} |= \mathbf{1}$.

We may repeat the same construction starting with another subgroupoid $\mathbf{G}_{B}$ associated with the maximal set of compatible observables $\mathbf{B}$  obtaining a Hilbert space $\mathcal{H}_{B}$ and a representation of $\mathbb{C}[\mathbf{G}_{B}]$ in terms of linear operators on $\mathcal{H}_{B}$.
If the space of events $\Omega_{A}$ and $\Omega_{B}$ are assumed to have the same (finite) cardinality (i.e., A and B codify for the same  physical information on the system under consideration) we have that $\mathcal{H}_{A}$ and $\mathcal{H}_{B}$ are isomorphic as Hilbert spaces.
More interestingly, the 2-groupoid structure of Schwinger's 2-groupoid $\Gamma$ appears represented by the hand of the theory of transformations of Hilbert spaces.  Consider a transformation $\varphi (\mathbf{a},\mathbf{a}'; \mathbf{b},\mathbf{b}' )\colon M_\mathbf{A}(\mathbf{a}, \mathbf{a}')  \Rightarrow M_\mathbf{B}(\mathbf{b}, \mathbf{b}')$ sending the selective measurement $M_\mathbf{A}(\mathbf{a},\mathbf{a}')$ into the selective measurement $M_\mathbf{B}(\mathbf{b},\mathbf{b}')$.   Following the ideas above, the selective measurements $M_\mathbf{A}(\mathbf{a},\mathbf{a}')$ and $M_\mathbf{B}(\mathbf{b},\mathbf{b}')$ will be represented on the corresponding Hilbert spaces $\mathcal{H}_\mathbf{A}$ and $\mathcal{H}_\mathbf{B}$ supporting the fundamental representations of the groupoids $\mathbf{G}_{\mathbf{A}}$ and $\mathbf{G}_{\mathbf{B}}$ respectively.    
%Hence we may represent an element $\Phi$ on the groupoid algebra of the Schwinger's 2-groupoid (recall that $\boldsymbol{\Gamma}$ is itself a groupoid) as an operator acting on the linear space span by selective measurements $M(\mathbf{a}, \mathbf{a}')$ by means of its fundamental representation, that is we consider the Hilbert space generated by the 1-cells of the 2-groupoid as the support of the fundamental representation of the groupoid.  
Then, considering the vector $| A \rangle =  \sum_{\mathbf{a}, \mathbf{a}' = 1}^n A_{\mathbf{a}, \mathbf{a}'}  | M(\mathbf{a}, \mathbf{a}') \rangle$ in  the Hilbert space generated by $\mathbf{G}_\mathbf{A}$ and the element $\Phi =  \sum T_{\mathbf{a},\mathbf{a}'; \mathbf{b},\mathbf{b}'} \varphi (\mathbf{a},\mathbf{a}'; \mathbf{b},\mathbf{b}' )$, we get (all repeated indexes are summed):
\begin{eqnarray*}
\pi (\Phi) A & = & \sum T_{\mathbf{a},\mathbf{a}'; \mathbf{b},\mathbf{b}'}  
A_{\mathbf{c}, \mathbf{c}'} \, \,  \delta(\mathbf{a}, \mathbf{c})  \delta(\mathbf{a}', \mathbf{c}') \, \,  \pi (\varphi (\mathbf{a},\mathbf{a}'; \mathbf{b},\mathbf{b}' ))| M(\mathbf{c}, \mathbf{c}') \rangle  \\ & = & \sum T_{\mathbf{a},\mathbf{a}'; \mathbf{b},\mathbf{b}'}  
A_{\mathbf{a}, \mathbf{a}'} | M(\mathbf{b}, \mathbf{b}') \rangle \, .
\end{eqnarray*}
But, transformations $\varphi (\mathbf{a},\mathbf{a}'; \mathbf{b},\mathbf{b}' )$ are defined by pairs of SG-measurements, that is the basis for the groupoid algebra $\mathbb{C}[\boldsymbol{\Gamma}]$ consists on pairs of SG-measurements, thus as a linear space it is the tensor product of the linear space generated by SG-measurements.   Then, the coefficients $T_{\mathbf{a},\mathbf{a}'; \mathbf{b},\mathbf{b}'}$ can be written as the products $T_{\mathbf{b}, \mathbf{a}} T_{\mathbf{a}', \mathbf{b}'}$, and we conclude:
\begin{equation}\label{matrix_transform}
\pi (\Phi) A  =  \sum T_{\mathbf{b}, \mathbf{a}} T_{\mathbf{a}', \mathbf{b}'} 
A_{\mathbf{a}, \mathbf{a}'} | M(\mathbf{b}, \mathbf{b}') \rangle  = \sum (T^\dagger A T')_{\mathbf{b}, \mathbf{b}'} | M(\mathbf{b}, \mathbf{b}') \rangle\, .
\end{equation}
where $T^\dagger AT'$ stands for the standard matrix multiplication of the matrices defined by $T = \left(\bar{T}_{\mathbf{a}, \mathbf{b}}\right)$,  $T' =  \left(T_{\mathbf{a}', \mathbf{b}'}\right)$ and $A = \left( A_{\mathbf{a}, \mathbf{a}'} \right)$.  The previous formula (\ref{matrix_transform}) shows that the transformation $\varphi (\mathbf{a},\mathbf{a}'; \mathbf{b},\mathbf{b}' )$ is represented in the standard form as an operator between the Hilbert spaces $\mathcal{H}_\mathbf{A}$ and $\mathcal{H}_\mathbf{B}$.

%%%%%%%%%%%%%%%%%%%%%%%%%%%%%%%%%%
%%%%%%%%%%%%%%%%%%%%%%%%%%%%%%%%%%

\section{Conclusions and discussion}

A careful analysis of the structure of the algebra of measurements  of a quantum system proposed by J. Schwinger reveals that its underlying mathematical structure is that of a 2-groupoid.  Using this background, a proposal for the mathematical description of quantum systems based on the primitive notions of \textit{events} or \textit{outcomes}, \textit{transitions} and \textit{transformations} under the mathematical form of a 2-groupoid is stated.   

The standard interpretation in terms of vectors and operators in Hilbert spaces is recovered when we consider the fundamental representation of such 2-groupoid.    Other representations can be chosen that will reveal different characteristics of the system.   

The analysis of the dynamics as well as other aspects of the theory like the reconstruction of the algebra of observables and the states of the quantum system from the 2-groupoid structure, and the statistical interpretation of the theory, have been barely touched and a detailed analysis   will be developed in forthcoming works.  It suffices to mention here that the algebra of the groupoid, which naturally has the structure of an involution algebra, gives a first insight regarding the connection with the $C^{*}$-algebraic formulation of quantum theories which will be further explored in forthcoming papers.
In particular, the projectors $|\mathbf{a} \rangle \langle \mathbf{a} |$  will turn out to be normal pure states of the $C^{*}$-algebra of the system, and the GNS construction applied to any of them will reconstruct the Hilbert space of the fundamental representation.   

Composition of systems, symmetries, significant examples and applications begining with the harmonic oscillator and other basic examples of quantum systems, etc., are all aspects that will be discussed in subsequent works.

%%%%%%%%%%%%%%%%%%%%%%%%%%%%%%%%%%
%%%%%%%%%%%%%%%%%%%%%%%%%%%%%%%%%%

\section*{Acknowledgments}

The authors acknowledge financial support from the Spanish Ministry of Economy and Competitiveness, through the Severo Ochoa Programme for Centres of Excellence in RD (SEV-2015/0554).
A.I. would like to thank partial support provided by the MINECO research project  MTM2017-84098-P  and QUITEMAD+, S2013/ICE-2801.   G.M. would like to thank partial financial support provided by the Santander/UC3M Excellence  Chair Program 2019-2020.
G.M. is a  member of the Gruppo Nazionale di Fisica Matematica
(INDAM),Italy.

%%%%%%%%%%%%%%%%%%
%%%%%%%%%%%%%%%%%%

\end{document}